\documentclass{emulateapj}
\usepackage{apjfonts}
\usepackage{natbib}
\usepackage{graphicx}
\usepackage{amssymb,amsmath}
\usepackage{xcolor}

\usepackage[citebordercolor=green]{hyperref}
\shorttitle{Improved ILC Method and CMB Angular Power Spectrum Estimation}

\begin{document}

\title{An Improved Diffuse Foreground Subtraction by ILC method: CMB Map and Angular Power Spectrum using Planck and WMAP Observations}

\author{Vipin Sudevan\altaffilmark{1}, Pavan K. Aluri\altaffilmark{2,3}, Sarvesh Kumar Yadav\altaffilmark{1},  Rajib Saha\altaffilmark{1}, Tarun Souradeep\altaffilmark{2}}

\altaffiltext{1}{Physics Department, Indian Institute of Science 
Education and Research Bhopal,  Bhopal, M.P, 462023, India.} 
\altaffiltext{2}{Inter-University Centre for Astronomy and Astrophysics, 
Post Bag 4, Ganeshkhind,   Pune 411007, India.} 
\altaffiltext{3}{Cosmology \& Gravity Group, Department of Mathematics and Applied Mathematics, University of Cape Town, Rondebosch 7701,
South Africa
}

\begin{abstract}
We report an improved technique for diffuse foreground minimization from Cosmic Microwave Background (CMB) maps
using a new {\it multi-phase iterative} internal-linear-combination (ILC) approach in harmonic space. 
The  new procedure consists of two phases. In phase 1, a diffuse foreground cleaned map 
is obtained by performing a  usual ILC operation in the harmonic space in a single iteration 
over the desired portion of the sky. In phase 2,  we obtain the final foreground cleaned map 
using an iterative ILC approach also in the harmonic space, however, now, during each iteration
of foreground minimization, some of the regions of the sky that are not being cleaned in the current 
iteration, are replaced by the corresponding 
cleaned portions of the phase 1 cleaned map. The new ILC method has an important advantage that
it nullifies a foreground leakage signal that is otherwise inevitably present in the old and usual 
harmonic space iterative ILC method. The new method is flexible enough to handle the input 
frequency maps, irrespective of whether or not they initially have the same instrumental and pixel 
resolution, by bringing them to a common and maximum possible beam and pixel resolution at the
beginning of the analysis, upto some suitably chosen different maximum values of multipoles 
for different input frequency maps. This dramatically reduces data redundancy and hence memory usage and 
computational cost during actual execution of the  foreground minimization  code. Even more importantly, 
during the ILC weight calculation it avoids any need to deconvolve partial sky spherical harmonic  coefficients by the beam 
and pixel window functions, which in strict mathematical sense, is not well-defined under the assumption 
of azimuthally symmetric window functions, which are applicable to only full sky spherical harmonic coefficients. 
Using WMAP 9-year and Planck-2015 published frequency maps we obtain a pair of foreground 
cleaned CMB maps with independent  detector noise properties. Each of these two foreground 
cleaned maps has higher instrumental resolution than the Planck-2015 published CMB maps. Using these cleaned maps we 
obtain CMB angular power spectrum for multipole range $2 \le \ell  \le 2500$. Our power 
spectrum matches well with the Planck  published results with some differences, at different multipole 
ranges.  We validate the entire procedure of the new technique by performing Monte-Carlo simulations. 
Finally, we show that the weights for ILC foreground minimization have an intrinsic 
characteristic that it tends to  produce a {\it statistically isotropic} CMB map as well.         
\end{abstract}

\keywords{cosmic background radiation --- cosmology: observations --- diffuse radiation}
\maketitle

\section{Introduction}
\label{Intro}
In recent years, after the observations of Planck~\citep{Planck2016,Planck2016lfi, Planck2016hfi, Planck2016_CMB} and 
WMAP~\citep{Hinshaw2013, Bennett2013} and other CMB experiments~\cite{ACT2013,SPT2_2014,SPT1_2013} 
Cosmic Microwave Background (CMB) radiation emerged as an important probe  to disinter a wealth of 
information about the physics of our Universe. The anisotropy field of CMB intensity Stokes parameter 
has not only become a probe to determine values of various cosmological parameters 
precisely~\citep{PlanckCosmoParam2016,Hinshaw2013}, but also it can constrain one of the fundamental concepts 
on which the current cosmology is based, the mechanism~\citep{PlanckInflation2016, WMAPInflation2003} 
of so-called  inflation~\citep{GuthInflation1981,StarobinskyInflation1982,LindeInflation1983}
- an epoch in very early universe which is defined by the stretching of any physical length scale in 
an increasing rate for some brief interval of time, as well as, properties related to the geometry and 
topology of the 3-dimensional space~\citep{Lachieze1995,Levin2002, Cornish2004, Bielewicz2012, Luminet2016, PlanckTopology2016}. 
The potential of CMB to gain deep understanding about the physical nature of our Universe is so enormous that it can 
even be used to shade some light on the very basic question: {\it Is different direction in space identical 
to each other?}~\citep{Copi2004,Eriksen2004, Eriksen2007, Samal2008,PlanckIso2016}. Further, it has been argued that the anisotropy 
in the CMB polarization Stokes (Q and U) signal can be used to investigate physical problems that are complementary 
to temperature only case~\citep{PolCom1993,PolCom1995,PolCom1997}, giving rise to additional information. Given the 
central role the CMB plays to decode the physics of our Universe, it is important to 
estimate the CMB signal by different independent research groups by employing independent statistical methods. In this 
work we desire to estimate a foreground cleaned CMB map and its angular power spectrum using observations from the WMAP 9-year and 
Planck-2015 data release.

One of the important methods to isolate a cleaned signal of CMB anisotropy from the foreground contamination 
is the so-called internal linear combination (ILC) approach~\citep{Tegmark96, Tegmark2003, Bennett03_fg, 
Saha2006, Saha2008,Saha2011, Saha2016}. Instead of using usual standard deviation as the measure of contamination 
~\cite{Saha2011} use a measure of non-Gaussianity given by sample kurtosis to estimate a foreground cleaned CMB map 
and its power spectrum from WMAP observations. The ILC method is unique in that it does not require 
to model the foreground components in terms of any templates and relies only upon the assumption that the 
distribution of CMB photon follows a blackbody distribution along every direction of the  sky, so that its 
temperature (anisotropy) is independent on the frequency. The method uses the principle of simple linear 
superposition of different frequency maps with certain amplitude terms associated with each frequency (called the weight 
factors). The ILC method has been used in CMB signal processing extensively for foreground subtraction and hence 
power spectrum estimation and  reconstruction of various foreground components. In  some recent publications some of the 
authors of this paper have studied the bias properties of ILC (in harmonic space) power spectrum  extensively~\citep{Saha2008} and 
extended this method to jointly estimate the spectral properties of foregrounds, their templates along with the 
CMB signal using simulated observations of low resolution polarization Stokes Q maps  of Planck and WMAP missions~\citep{Saha2016}.

In the usual ILC method to reconstruct the CMB signal, when performed in the spherical harmonic space in an iterative 
fashion, wherein each iteration is carried out to clean a given region of the sky, takes place a leakage of foreground 
signal from the region of the sky that are not cleaned yet, into the regions that is being currently cleaned at any 
given iteration. The leakage signal, if unaccounted for, leads to biased reconstruction of CMB signal. In this work, we improve the usual  
ILC approach by employing a technique that prevents the leakage phenomena completely. We further extend the usual ILC method
so that during the iterative ILC technique in the multipole space, one does not require to deconvolve the partial 
sky power spectrum by dividing them by the Legendre transforms of the azimuthally symmetric beam functions of different 
frequency maps, which strictly speaking, is not mathematically well-defined in a rigorous  sense. The method is also 
flexible enough, so that if one desires to mask off certain 
region of the sky which may be strongly contaminated, one can do so at the onset of the foreground removal. Masking the  
contaminated regions to begin with leads to important advantage that the ILC weights are not affected by the contaminated 
regions and provides a better foreground minimized CMB map for the rest part of the sky.      

There have been many attempts in the literature to estimate a  foreground cleaned image of CMB fluctuations.\cite{Bunn1994} 
and\cite{Francois1999a} develop a Weiner filtering approach.~\cite{Bennett03_fg, Hinshaw_07, Gold2009, Gold2011} use a template 
cleaning approach. A Markhov-Chain-Monte-Carlo based  approach was used for all component reconstruction by~\cite{Gold2009, Gold2011}. 
Reconstruction of CMB maps along with all other foreground components using Gibbs sampling approach have been 
implemented by~\cite{EriksenWMAP2007, Eriksen2008a, Eriksen2008b}.
~\cite{Delabrouille2009} use a needlet space ILC approach for reconstruction of CMB map.  

We organize our paper as follows. In Section~\ref{Problem} we define the basic problem and explain the motivation of 
the work in Section~\ref{Motivation}. We briefly review the old iterative ILC algorithm in Section~\ref{usual_ilc}. 
We describe the problems with the old algorithm in Section~\ref{Problems} and their remedy in Section~\ref{remedy}.
We describe new ILC algorithm which improves the old approach by preventing completely the foreground leakage in 
Section~\ref{new_ilc}. In Section~\ref{advantages} we highlight the advantages of the new ILC approach. In Section~\ref{isotropy}
we place the  usual ILC weights in the more general context of estimating a statistically isotropic CMB signal by minimizing 
some suitable combination of the Bipolar Spherical Harmonics~\citep{HajianSouradeep2003, HajianSouradeep2005,HajianSouradeep2006}, 
non-zero values of which are indicative of the breakdown 
of statistical isotropy. We describe the input frequency maps, beam window functions and point source mask file in Section~\ref{inputs}. 
In Section~\ref{Skyregions} we discuss method of dividing the sky into several regions, taking the intensity level 
of foreground emissions as in indicator for multiple  population of different foreground emitting sources. We describe the 
foreground removal and power spectrum estimation method in Section~\ref{method}. We discuss or results in Section~\ref{results}. 
In Section~\ref{simulations} we describe the simulation of the entire foreground removal and power spectrum estimation
method. Finally, we conclude and discuss in Section~\ref{D&C}.           
 
\section{Basic Problem}
\label{Problem}
In this work we focus on the problem of CMB map reconstruction  and CMB angular power spectrum estimation 
using the multi-frequency foreground (and detector noise) contaminated CMB maps as observed by WMAP and Planck 
satellite missions.  As discussed in Section~\ref{method}, we estimate CMB angular power spectrum using foreground 
minimized CMB maps by employing an improved and iterative ILC algorithm in the harmonic space. 

\section{Motivation}
\label{Motivation} 
The foremost motivation of this work is to reconstruction of maximum resolution (diffuse) foreground minimized 
CMB map and its power spectrum  using the foreground contaminated CMB maps of Planck and WMAP observations by 
employing an  improved version of ILC algorithm. Since as discussed in Section~\ref{Intro} both CMB maps and 
its angular power spectrum play crucial role in decoding physics of our universe it is utmost important to reconstruct 
these observables using multiple algorithms and by independent science teams. Our work, thus serve as a 
reconstruction of CMB map and its angular power spectrum independent of Planck science team. Overall, our reconstructed 
CMB map and its angular power spectrum match well the Planck  published results. 

\section{The old iterative ILC algorithm}
\label{usual_ilc}
\subsection{Linear Superposition of Data and Weights}
Let us assume we have full sky observations of foreground contaminated CMB maps  at  $n$ different 
frequency bands. Each of these maps is assumed to have same HEALPix\footnote{Hierarchical Equal Area 
iso-Latitude Pixelization on sphere, e.g., see~\cite{Gorski2005} for details.} pixel resolution  $N_{\textrm {side}}$ 
parameter.  At a given frequency band, $\nu_i$, where $i \in \{ 1, 2, ,3, ...n\}$, the net signal in 
harmonic space, for a given mode-index $(\ell, m)$, in thermodynamic temperature unit, may be written as, 
\begin{eqnarray}
a^i_{\ell m} = B^i_{\ell}P_{\ell}\left(a^c_{\ell m} + a^{f(i)}_{\ell m}\right) + a^{n(i)}_{\ell m}
\label{freq_map}
\end{eqnarray}
where $B^i_{\ell}$ and $P_{\ell}$ respectively denote circularly symmetric beam  and pixel window functions of the 
frequency map $i$. The pixel window function is independent on frequency $\nu_i$ since all input maps 
have same pixel resolution.  $a^c_{\ell m}$ denotes the CMB signal. $a^{f(i)}_{\ell m}$ and $a^{n(i)}_{\ell m}$ respectively
represent the net foreground and detector noise contamination at the frequency $\nu_i$.  

Using all the $n$ available frequency maps one defines a cleaned map in the harmonic space forming a 
linear superposition of all input frequency maps, 
\begin{eqnarray}
a^{\textrm {Cleaned}}_{\ell m} = \sum_{i=1}^n w^i_{\ell} \frac{B^0_{\ell}P_{\ell}}{B^i_{\ell}P_{\ell}}a^i_{\ell m}
\label{CMAP}
\end{eqnarray}
where  $B^0_{\ell}$ denotes beam window function of the highest resolution frequency map. We note that the 
pixel window function, $P_{\ell}$, cancels in the above equation since all the input maps are assumed to have 
same pixel resolution. The amplitude of superposition, $w^i_{\ell}$, denotes the weight factor for the 
frequency map $\nu_i$ for multipole $\ell$. The weights, at each multipole, for all frequency maps are obtained 
by minimizing the corresponding angular power spectrum, $\hat C^{\textrm {Cleaned}}_{\ell}$, of the cleaned map. 
As shown in~\cite{Saha2008} the choice of weights which minimizes the variance of the cleaned map at each 
multipole $\ell$, is given by, 
\begin{eqnarray}
{\bf W}_{\ell} = \frac{{\bf eC}^+_{\ell}}{{\bf eC}^+_{\ell}{\bf e}^T}
\label{weights}
\end{eqnarray}
where ${\bf W}_{\ell} = \{w^1_{\ell}, w^2_{\ell}, w^3_{\ell}, ...., w^n_{\ell}\}$ is a $1 \times n$,  row vector 
containing the amplitude of superposition for all frequency bands, ${\bf e} = \{1, 1, 1, ...1\}$ is an $1\times n$ 
array denoting the shape vector of CMB in the thermodynamic temperature unit,$\ ^+$ denotes the Moore-Penrose inverse,  
${\bf C}_{\ell}$ denotes an $n \times n$, cross-power (covariance)  matrix of the $n$ input frequency maps for multipole 
$\ell$. $(i,j)$th element of the covariance matrix is given by, 
\begin{eqnarray}
C^{ij}_{\ell} = \frac{{(B^0_{\ell})}^2}{B^i_{\ell}B^j_{\ell}}\sum_{m=-l}^{\ell}\frac{a^i_{\ell m}a^{j*}_{\ell m}}{2\ell+1}
\label{cross_cov}
\end{eqnarray}    
The power spectrum of the cleaned map is given by, 
\begin{eqnarray}
C^{\textrm{Cleaned}}_{\ell} = \frac{1}{{\bf e C}^+_{\ell}{\bf e}^T}
\end{eqnarray}

\subsection{Iterations}
In practice, the foreground spectral properties (e.g., synchrotron  and thermal dust) depend on sky positions. 
The efficiency of ILC foreground removal increases if foreground removal is performed in iterative fashion
wherein each iteration cleans a given region of the sky. The iterative foreground removal method in the multipole space has been described 
in~\cite{Tegmark2003, Saha2008}. In this approach, one first divides the sky in a total of $n_r$ number of 
disjoint regions. Starting with  the initial set of $n$ foreground contaminated frequency maps (we call them initial maps) 
one then choses to perform foreground removal in a total of $n_r$ iterations, wherein at iteration $i$ the $i$th
region is cleaned.  At any given iteration, wherein one cleans any one of the $n_r$ regions, one performs following 
operations successively on the initial maps.
\begin{enumerate}
\item{Choose the sky region to be cleaned and find the partial sky $a_{lm}$s from this region for each of these 
maps. Use these partial sky $a_{lm}$ to obtain weights for linear combination using the partial sky cross power 
spectrum (covariance matrix).}
\item{Expand the full sky initial maps into $a_{lm}$ and linearly superpose them in harmonic space using the 
weights for the currently chosen sky region. Convert the resulting $a_{lm}$ to a full sky map which is best 
cleaned in the current region. }
\item{Replace the current region of $n$ initial maps by the corresponding region of the cleaned map obtained 
in step 2 above, after properly taking care of beam resolutions of different frequency bands. Use these 
$n$ partially  cleaned maps, as the initial maps for the next iteration. }
\end{enumerate}
At the end of $n_r$ iterations all $n_r$ regions becomes foreground cleaned and one obtains $n$ foreground 
cleaned maps for $n$ frequency bands. The highest resolution cleaned map is  the one corresponding to the highest
resolution frequency map.  

\section{Problems with Old Iterative ILC algorithm}
\label{Problems}
The usual iterative ILC algorithm as described in Section~\ref{usual_ilc} has two major problems.
\begin{enumerate}
\item First, the iterative ILC method may cause  a foreground  leakage from regions that 
are not cleaned yet, into the region that is currently being cleaned. Intuitively, the leakage occurs 
since the $a_{lm}$'s on the right hand side of Eqn.~\ref{CMAP} are global quantities and therefore 
can cause foreground leakage into the region that is being cleaned from the rest of uncleaned region. 
 
\item Secondly, division by the beam window function to deconvolve the effect of the instrumental beam (Eqn.~\ref{cross_cov}), 
assuming azimuthally symmetric beam functions in pixel space, is not a mathematically well defined operation 
for partial sky analysis, since the partial sky $a_{lm}$'s are coupled due to masking effect and Legendre transform 
of the beam functions are applicable only to the entire sky spherical harmonics. This causes  the weights as defined by 
Eqn.~\ref{weights} to be sub-optimal when estimating them from a given sky region in the iterative ILC
method. 
\end{enumerate}
We present below a rigorous analysis of above shortcomings and describe our modifications to the usual 
iterative ILC algorithm so as to overcome both of these two problems. We call the modified algorithm 
{\it new} iterative ILC algorithm. We first describe the second shortcoming since it is simpler than 
the first case. 

\section{Remedy of Old ILC Problems }
\label{remedy}
\subsection{Beam Deconvolution} 
\label{improve1}
Before we discuss the remedy of the second shortcoming above, we first generalize the usual ILC method to 
handle frequency maps, each of which, in general, has a different $N_{\textrm {side}}$ parameter. This 
can be achieved in the following way. Let the $i$th frequency map has pixel and beam window function 
respectively $P^i_{\ell}$ and $B^i_{\ell}$. Finite beam and pixel resolution of the input maps will allow to 
perform spherical harmonic transform of each map upto a certain maximum multipole $\ell^{(cut)i}_{max}$, for the 
$i$th frequency map. We first bring all the frequency maps to the common beam and pixel resolution 
of highest resolution frequency map following,
\begin{eqnarray}
a^{0i}_{\ell m} = \frac{B^0_{\ell }P^0_{\ell }}{B^i_{\ell }P_{\ell }}a^i_{\ell m} 
\label{general_ilc}
\end{eqnarray}   
where $a^i_{\ell m}$ represents the spherical harmonic transform of input map of pixel resolution  $N^i_{\textrm side}$
and is given by Eqn~\ref{freq_map}. $P^0_{\ell}$ represents the pixel window function of largest pixel 
resolution map and $B^0_{\ell}$ represents the beam window function of the highest resolution frequency map. We 
convert spherical harmonic coefficients of the left hand side of above equation to the 
maps with $N_{\textrm side} = 2048$, which is the largest pixel resolution parameter of Planck and WMAP 
maps. In particular, during the (inverse) spherical harmonic transformation for frequency $\nu_i$ we use 
spherical harmonic coefficients  upto $\ell^{(cut)i}_{max}$ (e.g., see Table~\ref{ListMaps}). At this point, when all the input frequency maps 
have the same beam and pixel resolution functions, the usual ILC method has been rendered {\it general enough} 
to handle original frequency maps that  had different pixel resolution parameter~\footnote{We note that if one 
is interested in an ILC analysis in low-pixel resolution, one can as well chose a lower $N_{\textrm {side}}$ parameter
instead of the largest one, and corresponding pixel window function $P^0_{\ell}$ in Eqn~\ref{general_ilc}.}.         

Once we obtain the sets of $a^{0i}_{\ell m}$, which have same same beam and pixel resolution for different 
input frequency maps, we form the cleaned map following 
\begin{eqnarray}
a^{\textrm{Cleaned}}_{\ell m} = \sum_{i}^n w^i_{\ell } a^{0i}_{\ell m}
\label{cmap1}
\end{eqnarray}
We note that in Eqn~\ref{cmap1}, for any given $\ell$ the summation over $i$ runs only over those frequency maps for which 
the condition, $\ell \le \ell^{(cut)i}_{max}$, is satisfied. One can find the cleaned CMB map as given by  Eqn~\ref{cmap1} using the weights 
following Eqn~\ref{weights}, where the  elements of cross-power covariance matrix are given by, 
\begin{eqnarray}
C^{ij}_{\ell} = \sum_{m=-\ell}^l\frac{a^{0i}_{\ell m}a^{0j*}_{\ell m}}{2\ell +1}
\label{cross_cov1}
\end{eqnarray} 

{\it How does our new ILC method help remedy the second shortcoming?} We notice that it does so since the weights that 
appear in Eqn~\ref{cmap1} can now  be obtained from Eqn~\ref{weights} using the cross-power covariance matrix of 
Eqn~\ref{cross_cov1} which does not require any division by the beam window function. This is an important 
improvement over the usual iterative ILC method, since the weights determined at each individual multipole $\ell$ has now 
a well defined meaning associated with it.   

Apart from the remedy to aforementioned shortcoming,  bringing all maps to a common resolution at the beginning, has an 
additional advantage that any strongly foreground contaminated region may now be masked out, if desired so, 
before the foreground minimization starts. This  has an important implication in our analysis since this enables us to 
mask out  the localized resolved point sources  at the beginning of our high resolution analysis, which leads to better 
efficiency in removing defused foreground minimization. 

\subsection{Preventing foreground leakage}
\label{leakage}
In this section we aim to quantify the foreground leakage analytically. To better understand the the leakage 
phenomena we first consider a simple case and discuss the origin  of leakage. This helps us to identify the 
leakage term. We then investigate the leakage that occurs in usual iterative ILC foreground removal method 
and propose a methodology to nullify it.   
\subsubsection{A Simple Case}
\label{leak_simple}
\begin{figure*}[t]
 \includegraphics[scale=0.34,angle=90]{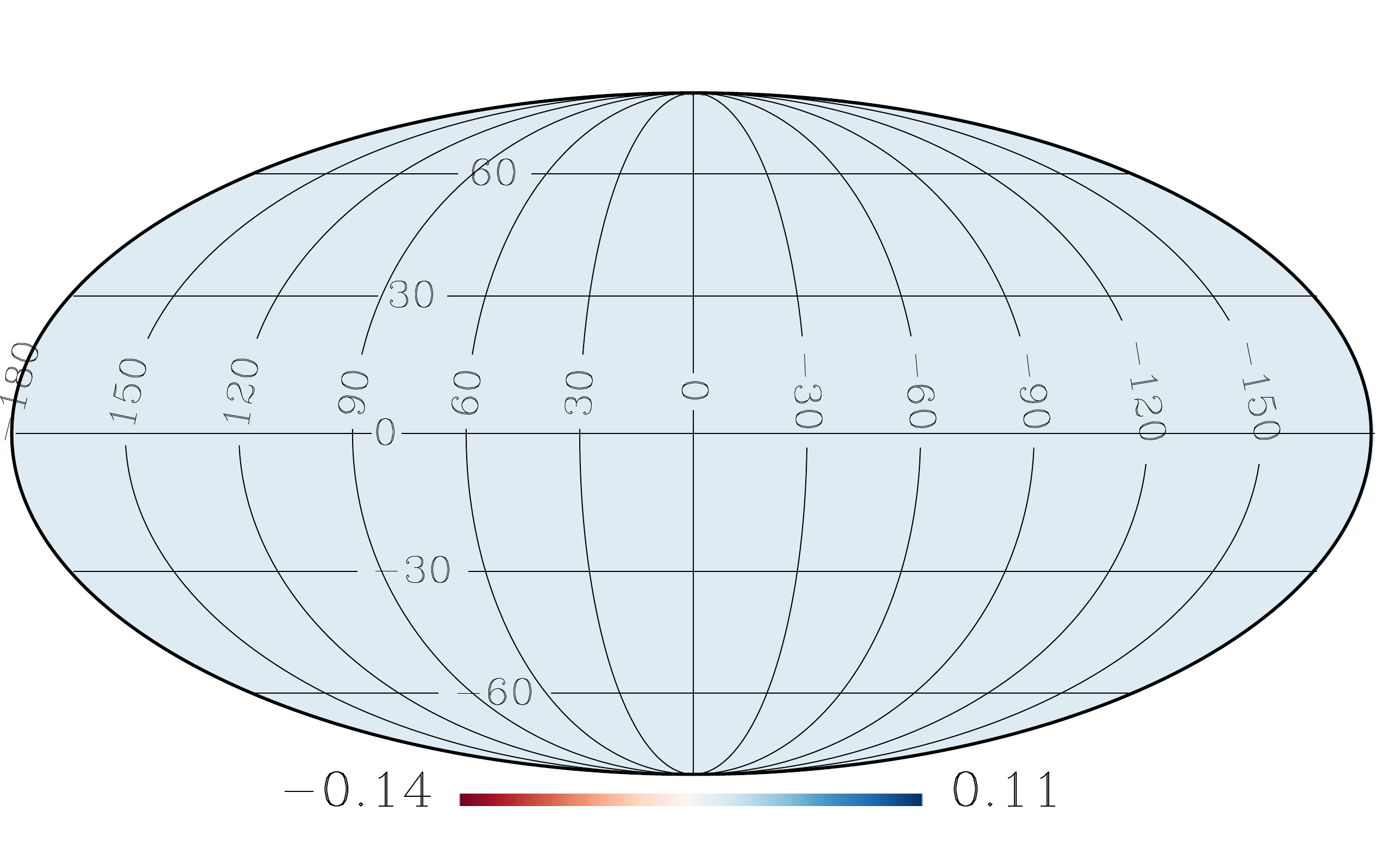}
 \includegraphics[scale=0.34, angle=90]{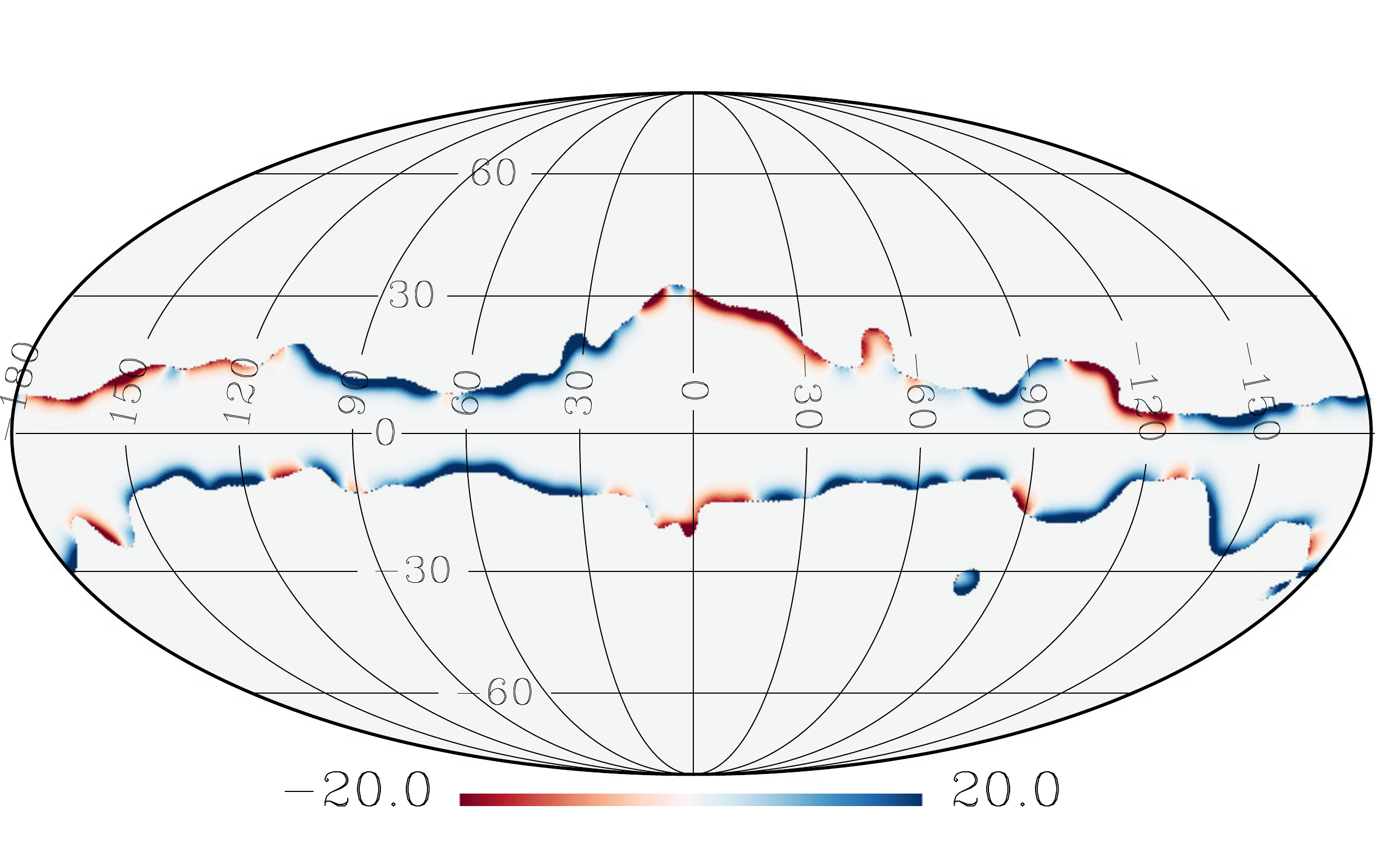}
\caption{Left: The difference between $f$ and $f'$ as mentioned in Section~\ref{leak_simple} in $\mu K$ 
thermodynamic unit. The difference map is consistent with zero, except near the locations of poles, where 
minor difference is expected to occur due to smaller number of pixels in the corresponding HEALPix rings. 
Right: The leakage  signal from region 2 into region 1 due to smoothing of spherical harmonic 
coefficients of CMB map after applying mask corresponding to region 1. Region 1 is defined by the area
from the galactic plane to the outer boundary formed by the non-zero pixels of this map, on 
both hemispheres. Region 2 is complementary to region 1. } 
 \label{leak1}
\end{figure*}

Let us consider a subdivision of the  entire sky into two disjoint regions, region 1 and region 2. The sky signal 
can be written as $f\theta, \phi) = f_1(\theta, \phi) + f_2(\theta, \phi)$, where $f_i$ ($i = 1, 2$) is non-zero 
only for the region $i$ and zero otherwise. Such a decomposition is valid since the two regions are disjoint and 
covers the entire sky. We can  rewrite this equation in the harmonic space as, $a_{\ell m} = \tilde a^1_{\ell m} + \tilde a^2_{\ell m}$, 
where $\tilde a^1_{\ell m}$ and $\tilde a^2_{\ell m}$ are the spherical harmonic coefficients of $f_1$ and $f_2$ respectively. 
That this result in harmonic space is true can be verified in the following way. 
\begin{enumerate}
 \item First generate a pure CMB map $f(\theta, \phi)$ from Planck best fit LCDM model at $N_{\textrm side} = 2048$. 
 \item Mask $f$ by the two complementary masks and obtain $f_1$ and $f_2$. 
 \item Obtain $\tilde a^1_{\ell m}$ and $\tilde a^2_{\ell m}$ from $f_1$ and $f_2$ upto $\ell_{max} = 4096$.  
 \item Obtain  $a'_{\ell m} = \tilde a^1_{\ell m} + \tilde a^2_{\ell m}$ for all $\ell \le \ell_{max}$. 
 \item Convert the $a'_{\ell m}$ above to a pixel map $f'(\theta, \phi)$ again for $\ell \le \ell_{max}$. 
 \item Verify that $f = f'$ 
\end{enumerate}
We show the difference map, $f(\theta, \phi)-f'(\theta, \phi)$ in the left panel of Figure~\ref{leak1} which is consistent 
with zero everywhere except at the locations near the poles where spherical harmonic transformations becomes somewhat 
erroneous due to lesser number of HEALPix pixels on the HEALPix isolatitude rings.

To introduce the leakage now we follow steps 1 to 3 as previously. We now form $\tilde a^{2'}_{lm} = B_{\ell} \tilde a^2_{\ell m}$ where 
$B_{\ell}$ is some filter function, that may as well represent the weight factors  in the context of  ILC foreground removal algorithm. 
If we make a map of $\tilde a^{2'}_{\ell m}$ the resulting map will not only completely occupy the region 2, but also it will extend to region 1. 
This we call {\it leakage}. Depending upon the exact form of the filter function, $B_{\ell}$, the leakage may be  quite different on 
case to case basis. We show the leakage by first forming a map from the spherical harmonic coefficient  
$\tilde a^1_{\ell m} + B_{\ell} \tilde a^2_{\ell m}$, then  subtracting map $f$ from this map, and finally masking the difference map by 
the mask corresponding to region 1. Mathematically, the masked difference map in the multipole space can be modeled as 
$ \tilde a^1_{\ell m} + K^1_{\ell m \ell' m'}B_{\ell'}\tilde a^2_{\ell' m'}$, where a summation over the repeated indices $\ell',m'$ are implied. 
$K^1_{\ell m\ell'm'}$ represents the mode-mode coupling matrix corresponding to region 1, as defined 
in~\cite{Hivon2002} . It is easy to note that the term $ K^1_{\ell m \ell 'm'}B_{\ell'}\tilde a^2_{\ell' m'}$ represents the leakage that occurs 
from region 2 to region 1. A pictorial representation of the leakage map (from region 2 to region 1) is shown in 
right panel of Figure~\ref{leak1} for $B_{\ell}$'s computed from a Gaussian beam function of FWHM = 300'. 

\subsubsection{Leakage in old iterative ILC method}
How does foreground leakage occurs in the usual and old iterative ILC method? To understand this let $a^{Iij}_{\ell, m}$ denotes the
spherical harmonic coefficients of the initial map corresponding to frequency $\nu_j$, at the beginning of the iteration $i$.
The $a_{\ell m}$  of the full sky cleaned map obtained at the step 2 of usual ILC iteration $i$ is now given by, 
\begin{eqnarray}
a^{ci}_{\ell m} = \sum_{j=1}^n w^{ij}_la^{Iij}_{\ell m}
\label{fsky_cmap}
\end{eqnarray}
We note that $a^{Iij}_{\ell, m}$ coefficients on the right hand side are partially cleaned for $i \ge 2$ and totally 
uncleaned for $i=1$. The spherical harmonic coefficients on the left hand side are cleaner than those of the right hand side 
because of the cleaning operation of the current iteration. The partial sky cleaned spherical harmonic coefficients corresponding to the 
$i$th region are given by, 
\begin{eqnarray}
\tilde a^{ci}_{\ell m} = K^i_{\ell m \ell' m'}a^{ci}_{\ell' m'}
\label{partsky_cmap}
\end{eqnarray}
where $K^i_{\ell m \ell' m'}$ represents the mode-mode coupling matrix for the $i$th region and a summation over repeated indices 
are assumed. The spherical harmonic coefficients of the initial maps at the beginning of  iteration $i$ are composed of three
classes. First, the partial sky {\it cleaned } spherical harmonic coefficients from the already cleaned regions (counting a total
of $i-1$ of them). Second, the spherical harmonic coefficients $\tilde a^{ij}_{\ell m}$ of the region to be cleaned in the current 
iteration, and thirdly, the spherical harmonic coefficients of the rest of the regions together ($\tilde a^{rest,ij}_{\ell m}$) which  are 
to be cleaned in $(i+1)$th iteration onwards. Note that, uncleaned partial sky spherical harmonic coefficients depend upon the frequency index 
$j$, whereas the spherical harmonic coefficients corresponding to the cleaned region are independent it. Mathematically, spherical harmonic 
coefficients of initial maps at the beginning of iteration $i$ are given by,  
\begin{eqnarray}
a^{Iij}_{\ell, m} = \tilde a^{c1}_{\ell m} + \tilde a^{c2}_{\ell m} + ....+ 
\tilde a^{ci-1}_{\ell m} + \tilde a^{ij}_{\ell m} + \tilde a^{rest,ij}_{\ell m}
\label{alm_initial}
\end{eqnarray} 
where the  first $i-1$ terms of the right hand side represents the already cleaned spherical harmonic coefficients.  Using Eqn.~\ref{alm_initial}
in Eqn.~\ref{fsky_cmap} we obtain, 
\begin{eqnarray}
a^{ci}_{\ell, m} = \tilde a^{c1}_{\ell m} + \tilde a^{c2}_{\ell m} + ....+ \tilde a^{ci-1}_{\ell m} 
+ \sum_{j=1}^nw^{ij}_{\ell}\tilde a^{ij}_{\ell m} + \sum_{j=1}^nw^{ij}_{\ell}\tilde a^{rest,ij}_{\ell m}
\label{fsky_cmap1}
\end{eqnarray}
where we have used the fact $\sum_{j=1}^nw^{ij}_{\ell} = 1$ for each multipole $\ell$  for any region $i$. 
Using Eqn~\ref{fsky_cmap1} in Eqn~\ref{partsky_cmap} we obtain, 
\begin{eqnarray}
\tilde a^{ci}_{\ell m} = K^i_{\ell m \ell' m'}  \left( \sum_{j=1}^nw^{ij}_{\ell'}\tilde a^{ij}_{\ell' m'} 
+ \sum_{j=1}^nw^{ij}_{\ell'}\tilde a^{rest,ij}_{\ell'm'}\right)
\label{cleaned_alm_i}
\end{eqnarray}   
where summation over repeated indices $\ell', m'$ are assumed. The spherical harmonic coefficients on the first term inside the bracket
contains contribution from CMB ($\tilde a^{Ci}_{\ell m}$), and foreground  $\tilde a^{Fij}_{\ell m}$ components (and of course, detector 
noise also, which we ignore for the time being since we are primarily interested in leakage 
caused by foreground components), hence, $\tilde a^{ij}_{\ell m} = \tilde a^{Ci}_{\ell m} +  \tilde a^{Fij}_{\ell m}$. Assuming a perfect
estimation of weights from the $i$th regions so that the term $\sum_{j=1}^nw^{ij}_{\ell} \tilde a^{Fij}_{\ell m}$ can be ignored, we obtain from
Eqn~\ref{cleaned_alm_i}, 
\begin{eqnarray}
\tilde a^{ci}_{\ell m}  =  \tilde a^{Ci}_{\ell m} +  K^i_{\ell m \ell' m'}   \sum_{j=1}^nw^{ij}_{\ell'}\tilde a^{rest,ij}_{\ell'm'}
\label{cleaned_alm_i_1}
\end{eqnarray}  
Clearly the second term of the right hand side represents the leakage term. Considering Eqn~\ref{cleaned_alm_i_1} we immediately draw  following 
inferences. 
\begin{enumerate}
\item Leakage occurs only from all regions with indices $>i$ to the current region $i$ and no leakage occurs from regions $<i$ to the current region. 
\item At any given ILC iteration when we are cleaning $i$th region of the sky the leakage is induced only into the $i$th region and into no other region. 
\item After a region has been cleaned no modifications to this region is done at any subsequent stage of cleaning.
\item Although $\tilde a^{rest,ij}_{\ell m}$ contains both the CMB and foreground signal from the uncleaned region, due to blackbody nature of CMB on 
all sky regions, only the uncleaned foreground portion contribute to the leakage since 
$K^i_{\ell m \ell' m'} \left(\sum_{j=1}^nw^{ij}_{\ell'}\right) \tilde a^{C,rest,i}_{\ell'm'} = 0$ 
as the $i$th sky region and rest of the region to be cleaned after the $i$th iterations, are disjoint. 
Thus no CMB signal takes part in leakage phenomenon.  
\end{enumerate}
From the last conclusion above, one can make an important observation. If $\tilde a^{rest,ij}_{\ell m} = \tilde a^{rest,i}_{\ell m}$, 
i.e., if the partial sky spherical harmonics from the uncleaned region, are independent on frequency $\nu_j$ then  
$K^i_{\ell m \ell' m'} \left(\sum_{j=1}^nw^{ij}_{\ell'}\right) \tilde a^{rest,i}_{\ell'm'} = K^i_{\ell m \ell' m'} \tilde a^{rest,i}_{\ell'm'}= 0$, hence the 
{\it leakage will be completely nullified.} Based upon this observation we propose the new multi-phase iterative ILC algorithm 
be performed in two phases.

\section{The New Multi-Phase Iterative Algorithm}
\label{new_ilc}
\begin{enumerate}
\item {\it Phase 1:} Perform a single iteration ILC map reconstruction over the sky region desired to be cleaned in harmonic space, using the input frequency maps.  
\item {\it Phase 2:} Now perform the old iterative ILC foreground removal method on the same set of input frequency maps as in the first phase, with one modification
in step 2 (e.g., see steps for old ILC as mentioned in Section~\ref{usual_ilc}). Before obtaining spherical harmonics of the cleaned  map over full sky at any iteration, 
replace the uncleaned regions of each one of the initial maps
by the corresponding region of the single cleaned map obtained in phase 1 above. Since the cleaned map obtained in the first phase is independent on frequency,
 no foreground leakage will occur in the iterative cleaning during phase 2.  
\end{enumerate}

\section{Advantages of the new method}
\label{advantages}  
\begin{deluxetable*}{cccccccccccc}
\centering
\tabletypesize{\small}
\tablewidth{0pt}
\tablecaption{Specifications of Input Frequency Maps}
\startdata
\hline 
\hline\\ 
 Set  & K1   &  30 GHz & Q1 & W1  & W2  & V1&  70 GHz& 100 GHz& 143 GHz &  217 GHz  &  353 GHz   \\   
 $S_1$ &  & (27M, 27S)        &    &     &     &   &  (DS1)      &  (DS1) &  (DS1)    &  (1,2)    & (Detector 1) \\\\ 
 $N_{\textrm side}$ & 512 & 1024 & 512 & 512& 512& 512& 2048 & 2048& 2048 & 2048& 2048 \\ \\
$\ell^{(cut)}i_{max}$    & 290  &  490    & 540& 740 & 740 & 790& 2038  & 2440       & 3240    &   3990      &    3990     \\   
\hline\\ 
 Set  & Ka1   & 30 GHz & Q2 & W3  & W4  & V2&  70 GHz& 100 GHz& 143 GHz &  217 GHz  &  353 GHz   \\   
 $S_2$ &   & (28M, 28S)   &    &     &     &   &  (DS2)  &  (DS2) &  (DS2)    &  (3,4)    & (DS2) \\ \\
 $N_{\textrm side}$ & 512 & 1024 & 512 & 512& 512& 512& 2048 & 2048& 2048 & 2048& 2048 \\\\ 
$\ell^{(cut)i}_{max}$    & 390  &  490    & 540& 740 & 740 & 790& 2038  & 2440       & 3240    &   3990      &    3990   
\enddata
\tablecomments{Table showing Planck and WMAP  frequency maps used in our work along with  
their specifications. The top panel shows the specifications for the set $S_1$ and the bottom panel 
shows the same for the set $S_2$. DS1 and DS2 respectively represents detector set 1 and detector set 2 
for Planck frequency bands, wherever applicable. The second row of each panel shows the the HEALPix pixel 
resolution parameter $N_{\textrm side}$ of the native WMAP and Planck frequency maps . The third row  
of each panel shows maximum values of $\ell$, given by $\ell^{(cut)i}_{max}$ variable for different input frequency 
maps for set $S_1$ and $S_2$. }
\label{ListMaps}
\end{deluxetable*}

Why should we prefer the new iterative ILC method over the old one? The main benefit of the new method is that it stops 
foreground leakage completely. Moreover, in the new method we consider all input frequency maps on equal footing as far 
as resolution is concerned, although,
initially they might have different beam and pixel resolutions. To convert resolutions of different input frequency
maps to a common  resolution, we  first upgrade pixel resolution of all input frequency maps to the highest pixel resolution of 
$N_{\textrm {side}} = 2048$ and  then bring them to the same instrumental beam resolution of the highest resolution map. A detailed description of 
actual of doing this is described in Section~\ref{improve1}. Bringing all maps at the common beam and pixel resolution has the advantage that 
only one cleaned map at the highest beam resolution need to be formed during each iteration of phase 2. Working with a single 
pixel resolution also makes it possible to use only a single mask file at the chosen pixel resolution encoding information about 
different sky region. These result in reduction in net memory requirement and execution time of the code, which is extremely helpful 
for performing massive Monte-Carlo simulations of the method. Moreover, since all frequency maps are in the same beam and pixel 
resolution at the beginning of foreground cleaning  we can use  Eqn~\ref{cross_cov1} instead of Eqn~\ref{cross_cov} to estimate 
the cross-power covariance matrix in the expression of weights as given by Eqn.~\ref{weights}  for any region $i$ of the sky. 
This avoids any need for deconvolution of partial sky spherical harmonic coefficients by the beam window function. Finally, 
bringing all maps to a common resolution to begin with allows one to mask out positions of strongly contaminated regions, if desired. 
This results in better performance of foreground removal from the rest of regions since now the weights are not affected 
by the undesired strongly contaminated region.  

\section{Relationship of ILC weights and the concept of isotropy}
\label{isotropy}
Although, Eqn~\ref{weights} was derived (e.g., see~\cite{Saha2008}) requiring that the angular power spectrum of the cleaned map is 
minimized subject to the constraint that CMB is projected out completely, assuming its blackbody spectrum, 
in this section we show that the ILC weights, as given by Eqn~\ref{weights}, can also produce a CMB cleaned map 
based upon a more fundamental concept, namely that of statistical isotropy of the CMB temperature field, than the 
relatively simpler concept of minimizing the variance of the cleaned map at different angular scale. Although, it is a harder problem 
than simple variance measurement, to define any unique quantitative measure of statistical isotropy, a set of 
statistics measuring statistical anisotropy as given has been proposed  by~\cite{HajianSouradeep2003,HajianSouradeep2005,HajianSouradeep2006} using the 
Bipolar Spherical Harmonics. The authors of the above reference define a set of estimators of statistically anisotropy following 
\begin{eqnarray}
\hat \kappa_{L} = \sum_{\ell_1, \ell_2, M} \vert \tilde A^{L, M}_{\ell_1,\ell_2}\vert^2
\label{kappal}
\end{eqnarray} 
where 
\begin{eqnarray}
\tilde  A^{L, M}_{\ell_1,\ell_2} = \sum_{m_1, m_2}a_{\ell_1 m_1} a_{\ell_2, m_2} \mathcal C^{L, M}_{\ell_1, m_1, \ell_2, m_2}
\end{eqnarray} 
where $ \mathcal C^{L, M}_{\ell_1, m_1, \ell_2, m_2}$ represents the usual Clebsch-Gordon coefficients. 
Using Eqn~\ref{kappal} one can define a measure  of statistical anisotropy, $\hat {\mathcal N}$,  which 
can be described  by a single scalar number for the entire  map, following, 
\begin{eqnarray}
\hat {\mathcal N} \equiv \sum_L\hat \kappa_L 
\label{iso1}
\end{eqnarray} 
Using the orthogonality of the Clebsch-Gordan coefficients, one can show that,  
\begin{eqnarray}
\hat {\mathcal N} = \sum_{\ell_1, \ell_2}  \hat C_{\ell_1, \ell_2}  
\label{iso2}
\end{eqnarray}
where 
\begin{eqnarray}
\hat C_{\ell_1\ell_2}= \sum_{m_1,m_2}\frac{\vert a_{\ell_1 m_1}a^*_{\ell_2, m_2}\vert^2 }{(2\ell_1 + 1) \times (2\ell_2+1)} 
\label{iso3}
\end{eqnarray}
What is the choice of weights that minimizes the statistic defined in Eqn~\ref{iso1} estimated from the cleaned map 
which is defined by Eqn~\ref{cmap1}. Using Eqn~\ref{cmap1} and Eqn~\ref{iso3}  in Eqn~\ref{iso2} we obtain, 
\begin{eqnarray}
\mathcal N = \sum_{\ell_1, \ell_2, m_1, m_2,i,j,p,q}\frac{w^i_{\ell_1}w^j_{\ell_2}a^i_{\ell_1m_1}
a^{j*}_{\ell_2m_2}w^p_{\ell_1}w^q_{\ell_2}a^{p*}_{\ell_1m_1}a^q_{\ell_2m_2}}{(2\ell_1+1)(2\ell_2+1)}
\end{eqnarray}
After some algebra, above equation can be written as, 
\begin{eqnarray}
\mathcal N = \left ( \sum_{\ell}{\bf W}_{\ell}{\bf C}_{\ell} {\bf W}_{\ell}^T \right)^2 
\end{eqnarray}
where ${\bf C}_{\ell}$ denotes the cross-power covariance matrix as defined by Eqn~\ref{cross_cov1}. Clearly, 
minimizing our measure of statistical anisotropies, ($\mathcal N$), is equivalent to 
minimizing the positive definite terms ${\bf W}_{\ell}{\bf C}_{\ell} {\bf W}_{\ell}^T $, which is
nothing but the power spectrum of the cleaned map at a multipole $\ell$.   
Minimizing the angular power spectrum of the cleaned map at each multipole subject to the constraint 
CMB signal is preserved we find weights  that minimizes $\mathcal N$, is  given by the ILC weights 
as given by Eqn~\ref{weights}. 

\section{Input Data Set}
\label{inputs}
\subsection{Frequency Maps} 

\begin{figure*}[t]
\includegraphics[scale=0.6]{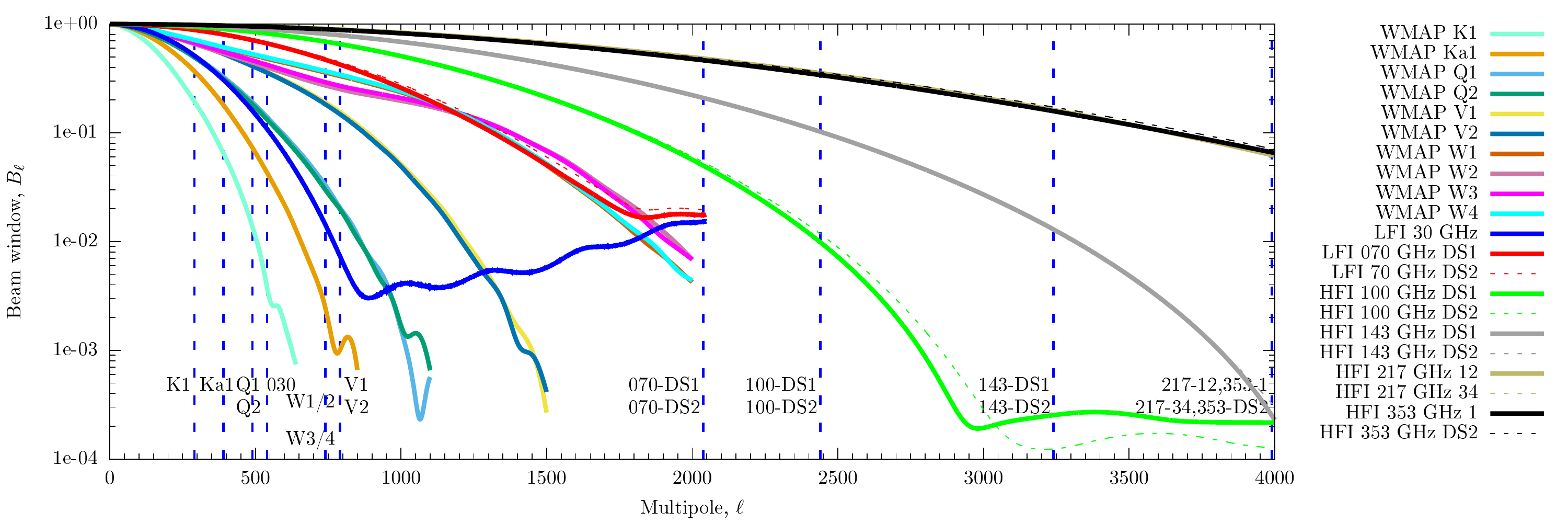}
\caption{Beam window functions, $B_\ell$, used in this work for WMAP and Planck frequency bands. The vertical 
dashed lines indicate the positions of the $\ell^{(cut)i}_{max}$ for any particular frequency map (see Table~\ref{ListMaps})
for details.}
\label{Bl}
\end{figure*}

We use a set of Planck LFI and HFI detector and  detector-set maps  and all 10 WMAP 
differencing assembly (DA) maps in our analysis. We do not use Planck 44 GHz frequency 
band since including maps of this frequency band we  recover unusually low CMB angular 
power spectrum, which possibly indicates presence of an unknown systematic effect (e.g., 
calibration issues) in the Planck 44 GHz frequency maps. Also, we do not use two 
highest frequency Planck HFI maps (e.g., 545 GHz and 857 GHz) since they are strongly 
contaminated by the thermal dust component. Since primary aim  of our current  work 
is to reconstruct improved foreground cleaned CMB signal though the iterative ILC method, 
 as well as estimation of CMB angular power spectrum, which we desire, should 
be free from detector noise bias, instead of producing merely a single foreground cleaned 
CMB map, we reconstruct two foreground cleaned CMB maps - each one of which has 
independent detector noise properties to the other. The  cleaned CMB maps with independent 
detector noise properties can then be used for CMB angular power spectrum estimation by a 
MASTER type~\citep{Hivon2002} cross-power spectrum estimation, which removes 
detector noise bias(e.g.,~\cite{Saha2006, Saha2008}). We form two cleaned maps with 
independent detector noise properties by using linear superposition within two disjoint sets 
of input frequency maps which do not contain any common detector or detector set map.
We label these sets as $S_1$ and $S_2$. We list all WMAP and Planck frequency maps used in 
our method in Table~\ref{ListMaps}. In the top and bottom panel of this table  
 we indicate whether the detector or  detector set map indicated by the first rows of the two panels 
belong to set $S_1$ or $S_2$.  For Planck 217 GHz we do not use the detector or detector set map. 
Instead we form input maps corresponding to $S_1$ and $S_2$ by averaging respectively detector 1,2 
and detector 3,4 maps within each pair.

\subsection{Beam Window Function}
For WMAP maps  we use WMAP 9-year published beam window functions corresponding to different DAs\footnote{available from 
https://lambda.gsfc.nasa.gov/product/map/dr5/beam\_xfer\_get.cfm.}. For Planck LFI frequency maps we use beam window functions 
for 30 GHz frequency map and individual beam window functions applicable for detector set 1 and detector set 2 for 70 GHz. 
LFI beam window functions were extracted from `LFI\_RIMO\_R2.50.fits' file. For HFI maps  we use beam window 
functions corresponding to different detector and detector set maps  as extracted from the file `HFI\_RIMO\_Beams-100pc\_R2.00.fits'.
Since for 217 GHz we average detector set 1, 2 and 3, 4 maps separately, we also average the beam window functions 
corresponding  to these detector maps for use with  the averaged maps. A plot of the beam window functions $B_{\ell}$ 
for different frequency maps of our analysis is shown in Figure~\ref{Bl}.  

\subsection{Point Source Mask}
\begin{figure}[htp]
\includegraphics[scale=0.35, angle=90]{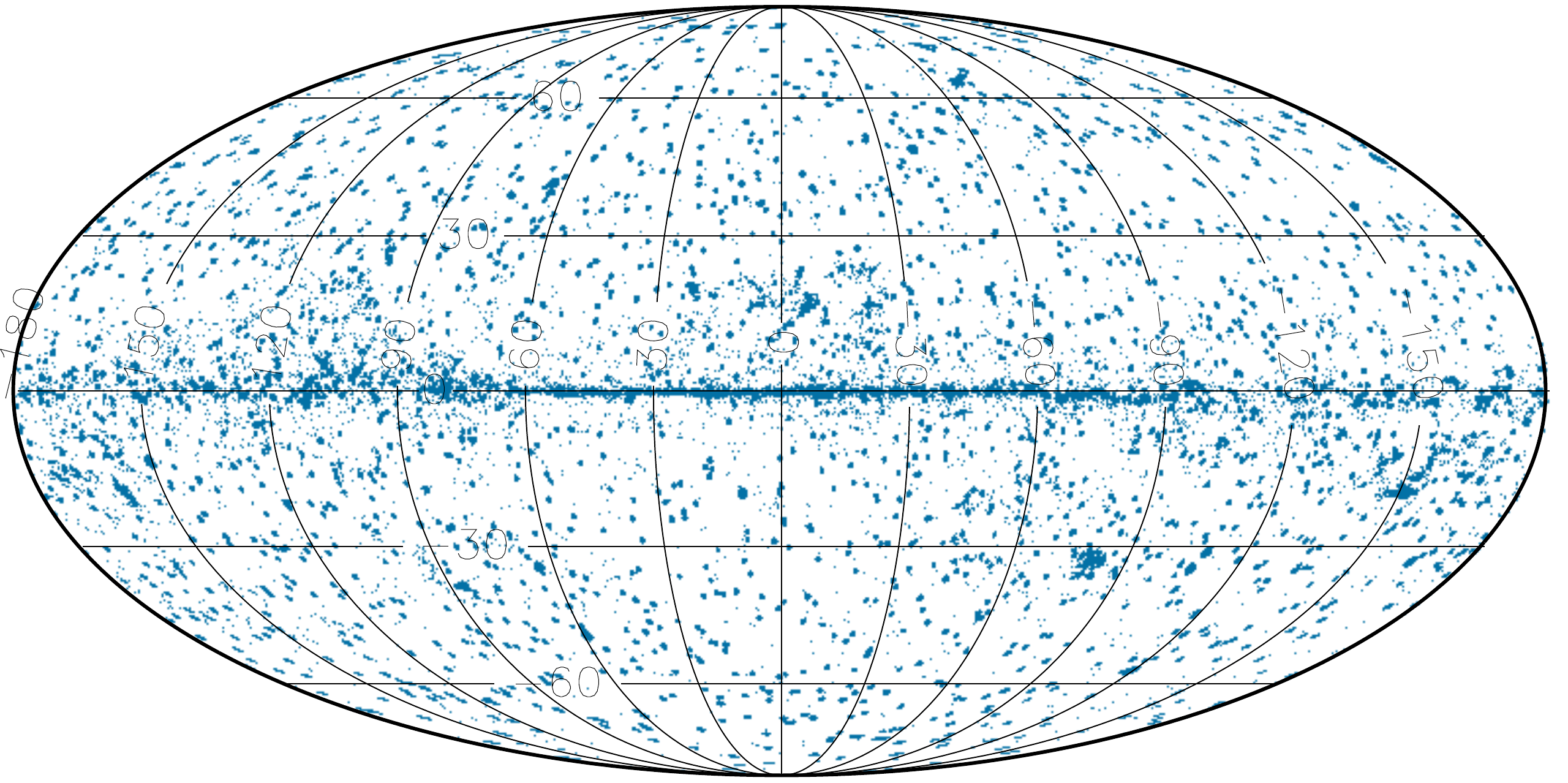}
\caption{Composite point source mask (PSMask) formed using the WMAP and Planck frequency maps 
in our analysis.}
\label{ps_mask}
\end{figure}
We use WMAP point source mask, Planck LFI point source mask, as well as Planck HFI point source 
masks at 100 GHz, 143 GHz, 217 GHz and 353 GHz frequency  bands to form a composite point source
mask for use in our analysis. We first upgrade the WMAP point source mask at $N_{side} = 2048$. 
We then multiply all masks at $N_{\textrm{side}} = 2048$ to form our composite point source mask 
(henceforth PSMask) at $N_{side} = 2048$.  The  `PSMask' covers  $89.5\%$ area of the entire sky.

\section{Masks that define the sky regions}
\label{Skyregions}
\subsection{Background}
Since the population of the foreground sources varies in their innate 
nature depending upon where they are located in our Galaxy (e.g.,  
galactic center, disk or halo region) ILC foreground removal method performs better 
when the sky containing the emission from the Milky Way is divided into 
a number of regions depending upon the nature of population of these 
sources. In earlier publications  some of the authors of this article~\citep{Saha2006,Saha2008}, 
presented an approach for sky divisions that is based upon the magnitudes 
of emission levels of foreground sources in Milky Way. 
As discussed in~\citep{Saha2008}, for an ideal experiment that is noiseless, 
all foregrounds can be completely removed if all of the following conditions 
can be met simultaneously: (1) foregrounds follow rigid frequency scaling all over the sky 
at the least over the entire frequency window of observation (2) total number of free 
parameters to  model the frequency variation of all foreground components 
are one less than the total number of detector maps available for linear 
combination and (3) any chance correlation between the CMB and other foreground 
components must be ignorable. Even if, there is only one foreground component 
all over the sky and  condition (1) and (3) are satisfied, a complete foreground 
removal will be impossible if the foreground component under consideration 
has some angular variation of its spectral index, e.g, synchrotron emission 
and thermal dust emission.  In this case, the foreground component with 
varying spectral index can be thought of a superposition of several foreground 
components each with a fixed and different spectral index all over the 
sky~\cite{Francois1999,Saha2016}. Such representation of a component with a 
varying index indicates that a foreground component with varying index can 
never be completely removed from the sky (on account of violation of condition 2 
above, even if one guarantees condition (1) and (3) are satisfied).

From the foregoing discussion one can infer that efficiency of ILC foreground removal
improves if one is able to find a region on the sky where different foreground 
components have approximately constant spectral indices. Finding sky regions with 
approximately constant spectral indices may appear relatively simpler if there were
only one foreground component on the sky. However, in presence of spectral index
variation of more than one components (e.g., synchrotron and thermal dust) a reliable
solution to this problem becomes almost impossible to attain, since different foreground 
components may have completely different morphological pattern of spectral index 
variation. Even in the case of spectral index variation due to only single 
component the variation may be on very small angular scale on the sky (e.g., two
different but spatially close by population of localized sources). In principle, an optimal 
choice of sky region on which the spectral index remains roughly constant for a single 
component may even become invisible without further zooming in the desired region of the sky. 
These requirements demand that one performs foreground removal in increasing number of sky divisions  
which increases total number of iterations required for foreground removal 
making the ILC method computationally very expensive.

Given the complexity described above in defining the sky regions with constant 
spectral indices we follow an approach which is similar to  foreground intensity 
based sky division approach as in~\cite{Saha2006, Saha2008}. 
In the following, we discuss the procedure for sky divisions based upon the intensity levels 
of net foreground emission.    

\subsection{Sky regions}
\begin{figure}
\includegraphics[scale=0.35]{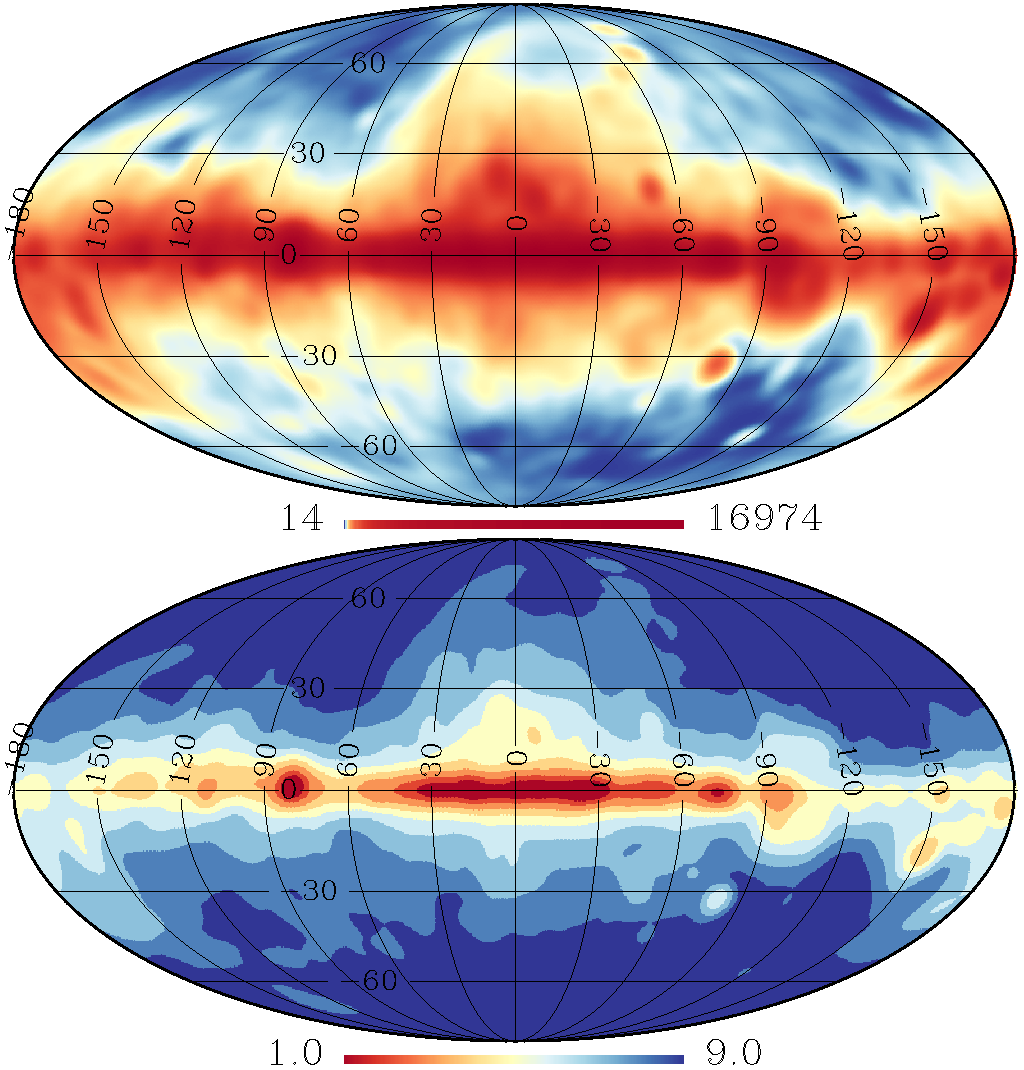}
\caption{{\it Top:} Figure showing the pixel temperature distribution (in $\mu K$ thermodynamic temperature 
unit) for low frequency foreground components. A histogram equalized color-scale is used to take 
into account the wide dynamical range of the image of this figure. {\it Bottom:} All the nine sky regions indexed 
from one to nine, depending upon the level of foreground contamination for high resolution analysis 
of this work.}
\label{intensity_mask}
\end{figure}

For our analysis at $N_{\textrm side} = 2048$, we make a mask that defines different sky regions which is 
determined primarily by the emission pattern of foreground components at the low frequency side of WMAP 
and Planck observation window. For this purpose, we take the Planck
70 GHz DS1 and DS2 maps and downgrade their pixel resolution to $N_{\textrm {side}} = 256$\footnote{We 
chose $70$ GHz since it is in overall the least foreground contaminated frequency
band for WMAP and Planck observation.}. 
We then smooth these maps by a Gaussian beam function of FWHM = $360^\prime$ in the harmonic
space after first deconvolving their spherical harmonic coefficients by their individual 
beam window function. We average these smoothed maps to form a single map at 70 GHz. In 
a similar fashion we downgrade the pixel resolution of  WMAP K1 band frequency map to 
$N_{\textrm {side}}= 256$ and smooth this map by a Gaussian  beam window of FWHM = $360^\prime$ 
after first  deconvolving by its original beam function. We subtract the 70 GHz smoothed averaged 
map from the K1 band smoothed map to form a map free from CMB signal and dominated by the 
low frequency foreground components. We show the resulting difference map in the top panel of 
Fig.~\ref{intensity_mask} and it has a maximum and minimum 
value respectively $16974.539$ and $14.454618 \mu K$. We make nine sky regions from this map 
with pixel temperature values, $\Delta T(p)$ lying in the following ranges: $ \Delta T(p) < 100$, 
$100 \le \Delta T(p) < 200$, $200 \le \Delta T(p) < 400$, $400 \le \Delta T(p) < 800$, 
$800 \le \Delta T(p) < 1600$, $1600 \le \Delta T(p) < 3200$, $3200 \le \Delta T(p) < 6400$, 
$6400 \le \Delta T(p) < 10000$  and finally $\Delta T(p) \le 10000 \mu K$. The resulting 
sky regions contains some isolated very small regions over different parts of the sky. In Table~\ref{SkyRegion1}
we reassociate these regions with the bigger sky regions that surrounds them keeping the 
total number of disjoint sky regions as nine. The resulting sky-region definition map contains pixel 
values from one to nine starting from the most contaminated region. Since our high resolution 
analysis were performed  at $N_{\textrm {side}} = 2048$ we upgrade the pixel resolution of the 
sky-region definition map to $N_{\textrm {side}} = 2048$. We show the sky regions  for high resolution 
analysis in the bottom panel of Fig.~\ref{intensity_mask}.

\begin{deluxetable}{cccc}
\tabletypesize{\small}
\tablewidth{0pt} 
\tablecaption{Definition of Very Small Sky Regions for Intensity based Sky Divisions}
\tablehead{\colhead{Sky location, $(\theta, \phi)$}          &    
           \colhead{Index for initial sky region }  &
           \colhead{Index for final sky region}     }   
\startdata
$55^\circ > \phi \ge 45^\circ$      &   2 & 3           \\  
$140^\circ > \phi \ge 130^\circ$    &   3 & 4          \\  
$300^\circ > \phi \ge 280^\circ$    &   7 & 8          \\   
$30^\circ \ge \theta \ge 20^\circ$  &     &           \\ 
$70^\circ > \phi \ge 60^\circ$      &  8  & 9          \\  
$60^\circ \ge \theta \ge 40^\circ$  &     &    
\enddata
\tablecomments{{\ \ List of very small sky patches which are redistributed to  neighboring
bigger sky region in intensity based sky division scheme. The first column shows the location of the 
sky patches in the usual spherical polar coordinate ($\theta, \phi$). The second column indicates the 
initial sky region to which the patch originally belonged to. The third column shows the sky region 
to which a small sky region with the corresponding specifications given in the first and second column 
is merged after the redistribution. }   
}  
\label{SkyRegion1}
\end{deluxetable}

\section{Methodology} 
\label{method}
\begin{figure*}[t]
\includegraphics[scale=0.45]{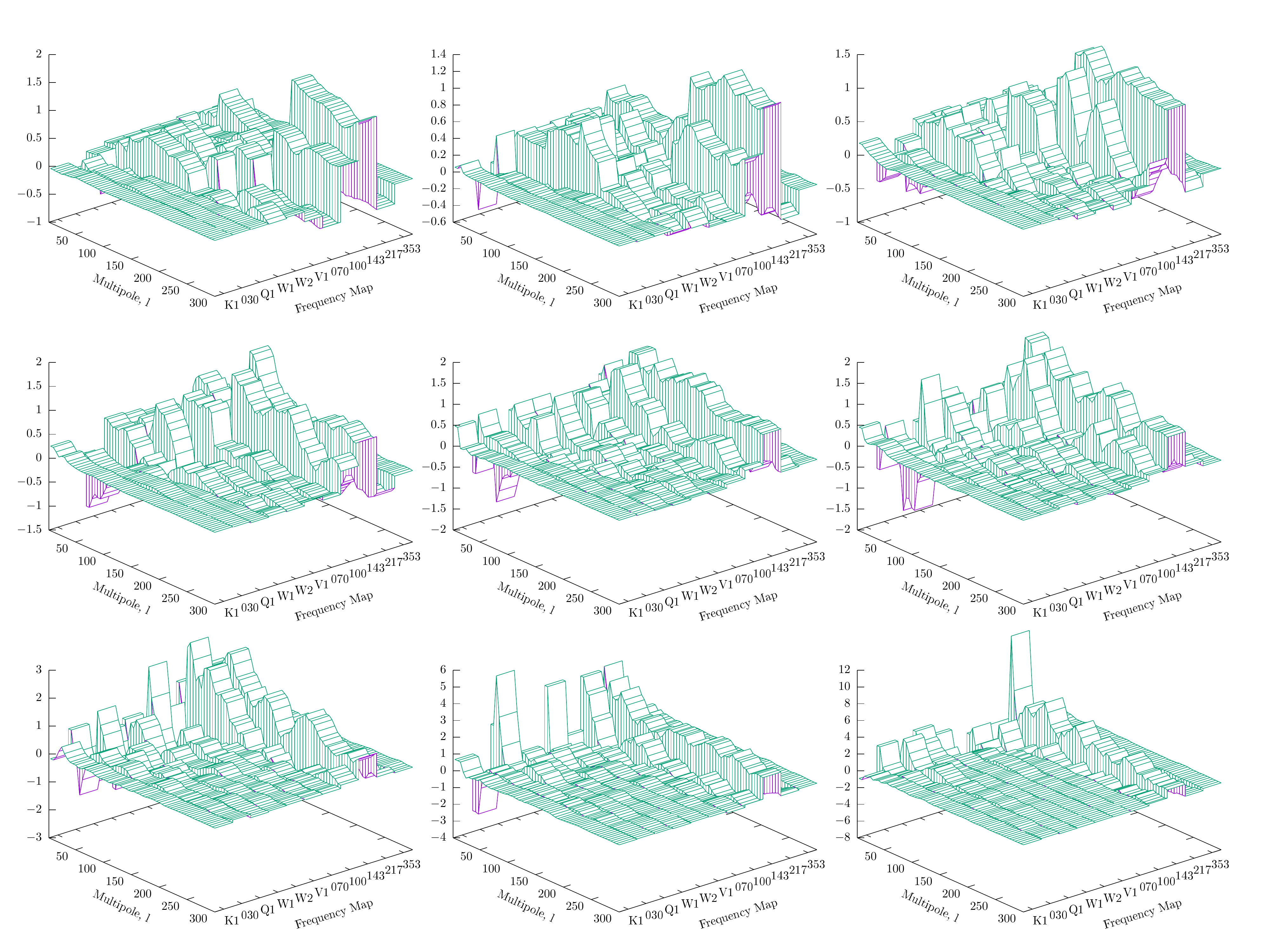}
\caption{From top to bottom and left to right weights are plotted for different sky regions for low multipole range. 
For visual purposes weights are plotted after every five multipoles.}
\label{wt_plot}
\end{figure*}

All the WMAP DA maps are provided by the WMAP science team in $mk$ thermodynamic temperature 
unit, whereas the Planck detector and detector set maps are provided in Kelvin thermodynamic 
temperature unit. We first convert all the input frequency maps as listed in table~\ref{ListMaps} 
in $\mu K$ (thermodynamic) temperature unit. As listed in this table, each of the 
two sets, $S_1$ and $S_2$, in our analysis, consists of a total of $11$ maps. The highest resolution 
frequency maps for set $S_1$ and $S_2$ are respectively detector 1 map  and detector set  2 map of 
$353$ GHz frequency band. 

We perform foreground minimization at the highest possible pixel resolution, $N_{\textrm side} = 2048$, of 
Planck and WMAP maps. For this purpose  we first expand all  input 
frequency maps of set $S_i$  where ($i=1,2$) in spherical harmonic coefficients and smooth them to the beam 
and pixel resolution ($N_{\textrm side} = 2048$) of the highest resolution map within each set by multiplying them 
by the ratio of the pixel times beam window function of the highest resolution frequency map within each set to the 
pixel times beam window functions of their native resolution, following Eqn~\ref{general_ilc}. Since the beam 
window functions of low frequency Planck  maps and WMAP maps  approaches 
zero faster than those of high resolution maps (e.g., see Fig.~\ref{Bl}) we apply a suitably chosen cutoff on 
the maximum $\ell$ ($\ell^{(cut)i}_{max}$) upto which spherical harmonic transformation is performed in Eqn~\ref{general_ilc}
for each input frequency maps.  The choice of these maximum $\ell$ values help  avoid division by numerically very 
small numbers in Eqn~\ref{general_ilc}. The $\ell^{(cut)i}_{max}$ values for different WMAP and Planck frequency maps 
are listed in Table~\ref{ListMaps}. We convert the  smoothed spherical harmonic coefficients to upgraded resolution 
input maps at $N_{\textrm side} = 2048$, upto $\ell^{(cut)i}_{max}$ values as listed in table~\ref{ListMaps}, by performing a 
backward spherical harmonic transformation using HEALPix supplied facility {\tt synfast}.  Once all maps have been 
upgraded in their beam and pixel resolution we apply our point source mask (PSMask) on each one of them to mask 
out locations of bright sources. Masking out the locations of bright sources at the very beginning of the analysis 
has an advantage  that the weights for different regions are not affected by these sources and ensures better performance 
for removal of diffuse foreground emission.   

Using point source masked frequency maps at $N_{side} = 2048$ we perform foreground removal in {\it phase 1} over 
the point source masked sky in a single iteration for both sets $S_1$ and $S_2$ and obtain two cleaned maps 
${\tt C_1P_1}$ and ${\tt C_2P_1}$.  In {\it phase 2}, we follow the new iterative ILC algorithm. Here we clean point source masked 
sky in a total of nine iterations over nine different sky regions (shown in Fig.~\ref{intensity_mask}) successively 
starting from the most contaminated region. As discussed in Section~\ref{Skyregions} different sky regions of 
Fig.~\ref{intensity_mask} indicate regions of the sky where the population of foreground are likely to differ resulting 
in different spectral behavior.  While estimating the spherical harmonic coefficients of the full-sky cleaned map  
at each iterations for set $S_1$ or $S_2$ in {\it phase 2} analysis, we replace the yet uncleaned regions of the partially cleaned frequency maps by 
the corresponding region of cleaned map {\tt CiP1} where $i=(1,2)$. As discussed in Section~\ref{leakage},  this replacement nullifies 
the foreground leakage from the uncleaned regions into the region that is being cleaned at any given iteration. For both 
phase 1 and phase 2 analysis we restrict any forward and backward spherical harmonic transformations 
upto $\ell^{(cut)i}_{max}$ values as specified in Table~\ref{ListMaps}. We denote the two cleaned maps corresponding to set $S_1$ and $S_2$, 
obtained at the end of {\it phase 2} analysis respectively by ${\tt C_1P_2}$ and ${\tt C_2P_2}$.      

How much improvements do we achieve in the new iterative ILC method over the old one? To see this we also perform foreground 
removal using frequency maps from sets $S_1$ and $S_2$ using the old ILC method. We call this {\it phase 3} analysis. Unlike, 
cleaned maps obtained from the {\it phase 2} analysis, the two cleaned maps ${\tt C_1P_3}$ and ${\tt C_2P_3}$ obtained from {\it phase 3}
analysis are independent on {\it phase 1} cleaned maps  ${\tt C_1P_1}$ and ${\tt C_2P_1}$. 

\section{Results from Planck and WMAP Observations}
\label{results} 
\subsection{Weights} 
How do the weights vary with sky regions and multipole, $\ell$? We show this variation in Figure~\ref{wt_plot} for the lower 
side ($2\le \ell \le 300$) of the entire range of multipoles  which are used to produce cleaned map $\tt C_1P_2$, obtained 
from set $S_1$ input map of the {\it phase 2} analysis. From the topmost left to the bottom right, the plots of this figure 
show the weights for the most to least contaminated regions of the sky. 
As we see from this figure, weights show different types of variation with respect to the multipole 
moments for each region, which indicates that the foreground spectra varies from region to region. 
At the high multipoles where only two Planck  frequency bands 217GHz and 353GHz  have the comparable resolution, all  weights tend to the 
Planck 217GHz frequency maps since it has the lower detector noise level between these bands. 

\subsection{Cleaned Maps}
\begin{figure}
\centering
\includegraphics[scale=0.48]{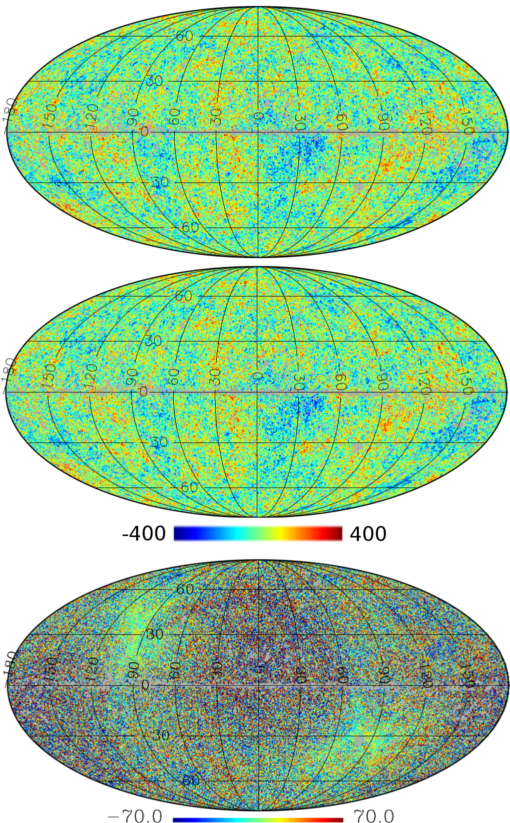}
\caption{{\it Top:} Cleaned map obtained from Planck and WMAP observed frequency maps corresponding to set $S_1$
using the new ILC algorithm described in this work. {\it Middle:} Cleaned map obtained from frequency maps of 
set $S_2$. Both these maps are plotted in a color scale in $\pm 400 \mu K$ range. {\it Bottom:} The difference between the top and middle
panel. This map is dominated by the detector noise. All maps plotted in thermodynamic temperature unit.}
\label{cmaps}
\end{figure}

We obtain two cleaned maps, ${\tt C_1P_2}$ and ${\tt C_2P_2}$, at the end of nine iterations of 
phase 2 analysis using the frequency maps from set $S_1$ and $S_2$  respectively. Since, we mask 
out positions of the known point sources by applying `PSMask' at the beginning of both phase 1 
and phase 2 foreground analysis,  our final cleaned maps  are also masked out at these positions. 
Both these maps are produced at the maximum resolution within a given set (i.e., $\tt C_1P_2$ has 
beam of detector 1 of 353 GHz frequency map, and  cleaned map $\tt C_2P_2$ has  the beam resolution 
of detector set 2 of 353 GHz frequency map). We have shown these cleaned maps for set $S_1$ and $S_2$ 
respectively,  at the top and middle panel of figure~\ref{cmaps}. In the bottom panel of the figure 
we have shown the difference  between the top and middle panel figures.  As clearly visible from this 
figure the difference map contain no visible signature of foreground emissions and 
dominated by the detector noise. This indicates that, both the maps are cleaned effectively. We note 
that resolution of cleaned map ${\tt C_2P_2}$ is slightly better than the resolution of cleaned map
${\tt C_1P_2}$ (e.g., see Fig~\ref{Bl}.). We therefore interpret the difference map primarily 
for the visual inspection of any residual foreground contamination left over in any one of the  two cleaned 
maps. The presence of  detector scan pattern as induced in the difference map shows that it is the 
detector noise and not the residual foregrounds that contributes dominantly to the residuals of the 
cleaned maps  ${\tt C_1P_2}$ and ${\tt C_2P_2}$. 

\begin{figure}
\includegraphics[scale=0.24]{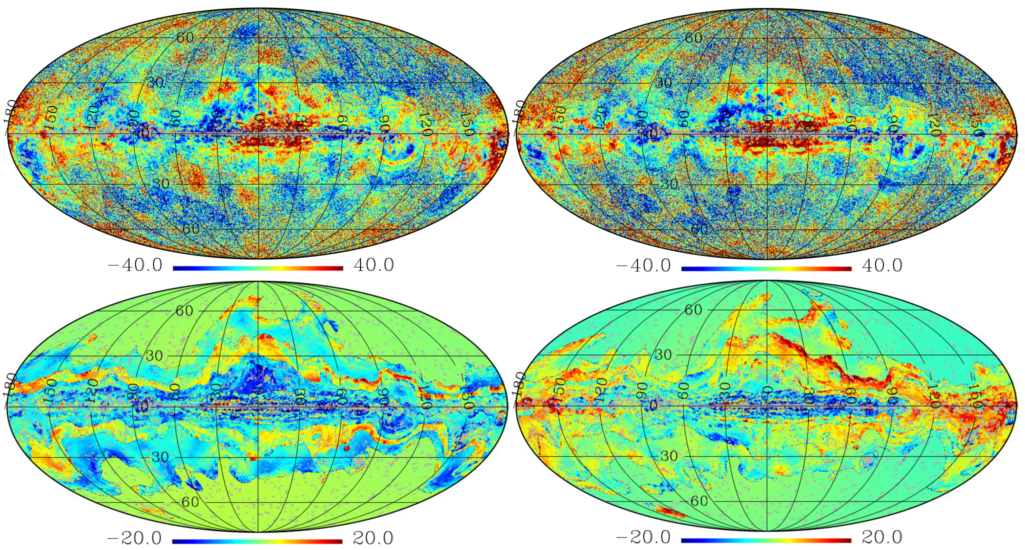}
\caption{{\it Left:} Top figure shows the difference between the cleaned maps obtained in phase 1 (single iteration)
and  phase 2 (i.e., following the new ILC algorithm developed in this work), both cleaned maps being estimated 
for set $S_1$ frequency maps of Planck and WMAP   observations. This figure primarily represents foreground residual 
that is present in single iteration cleaned map compared to the multiple iteration phase 2 cleaned map. 
Bottom figure shows the difference between the phase 3 and phase 2 cleaned maps.
The difference map shows the leakage present in the phase 3 cleaned map. {\it Right:} Same as left panel but for set $S_2$.  }
\label{DiffMaps}
\end{figure}

Since we obtain the cleaned map in phase 1 in a single iteration over the point source masked sky, it contains
some residual contamination compared to phase 2 cleaned map. We show the difference between the  phase 1 and 
phase 2 cleaned maps in the top panel of Figure~\ref{DiffMaps}. The left  figure of the top panel shows the 
difference between the cleaned maps for set $S_1$ (i.e, {$\tt C_1P_1-C_1P_2$}) whereas the right panel shows the difference between the cleaned maps 
for the set, $S_2$ (i.e, {$\tt C_2P_1-C_2P_2$}). The imperfect foreground removal of the phase 1 cleaned maps
compared to the phase 2 cleaned map is  clearly reflected in the both the difference maps of the  top panel 
of this figure. The bottom panel of figure~\ref{DiffMaps} shows the difference between the phase 3 and phase 2 
cleaned maps for set $S_1$ and $S_2$ respectively. The features of this map is dominated by the foreground 
leakage signal present in the phase 3 cleaned map.

\begin{figure}
\includegraphics[angle=90,scale=0.35]{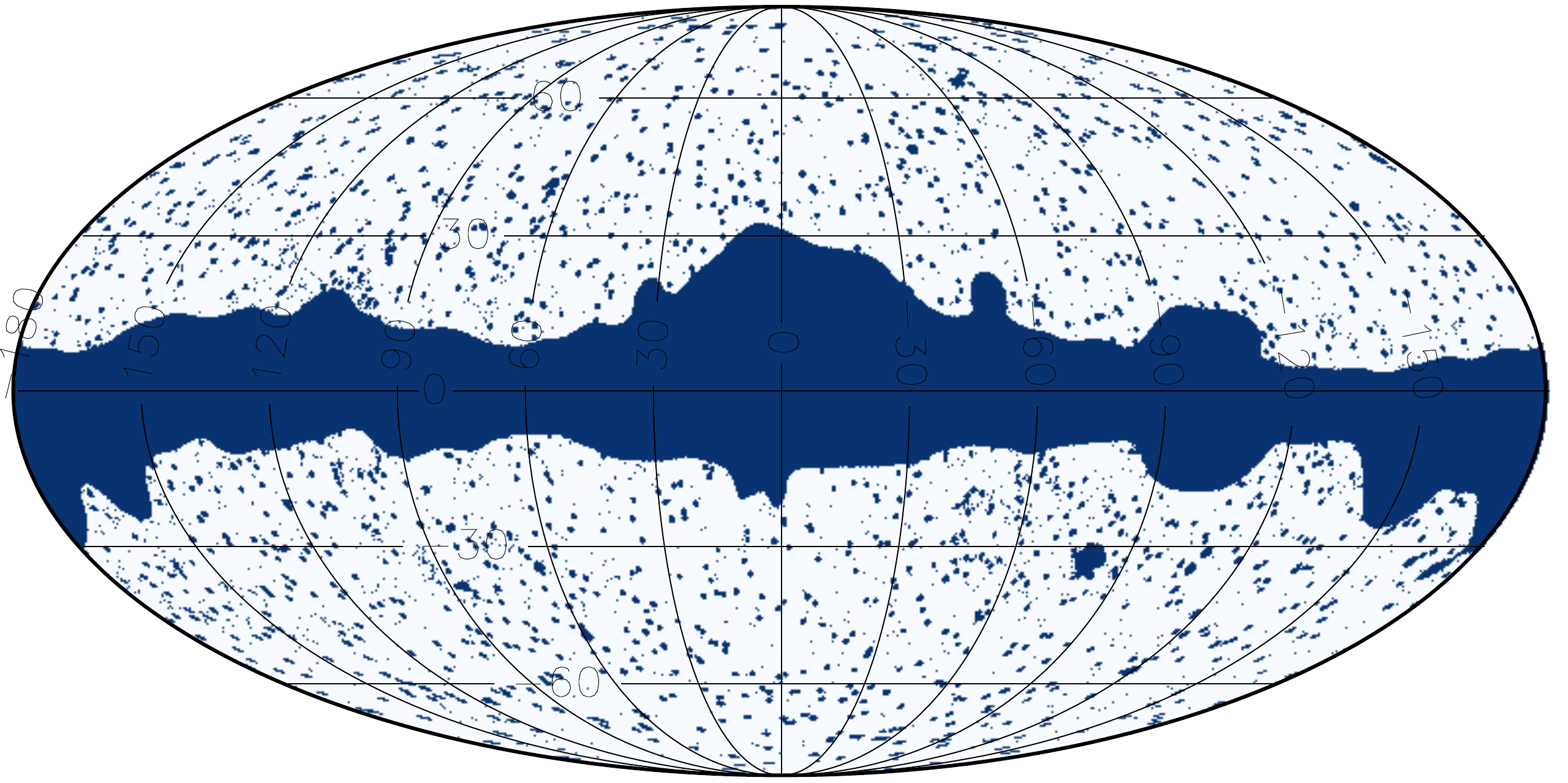}
\caption{M6 mask that excise pixels  from the first 6 most contaminated regions of Figure~\ref{intensity_mask} along with the  
locations of point sources as determined by the `PSMask'. White region of this plot shows the pixels that survive after masking.}
\label{M6Mask}
\end{figure} 

\subsection{Power Spectrum} 
\begin{figure*}
\includegraphics[scale=0.6]{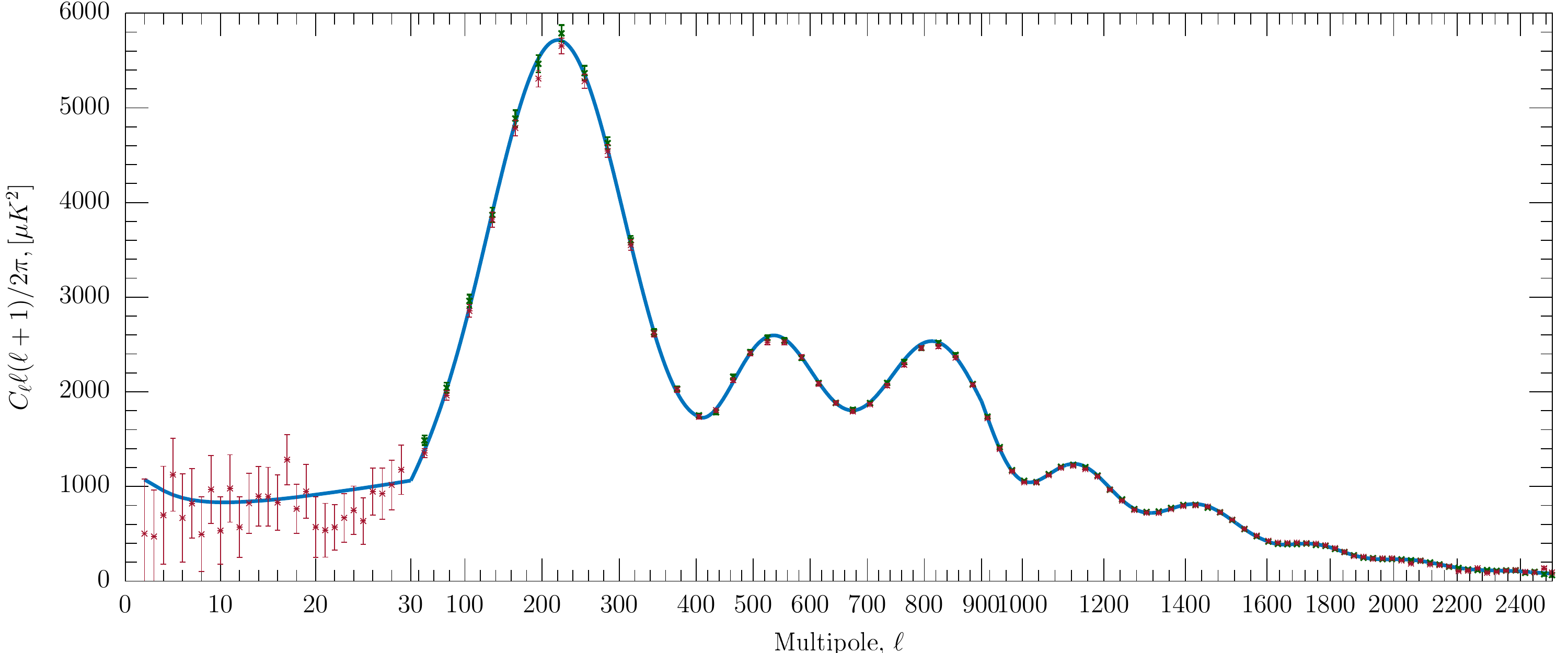}
\caption{CMB angular power spectrum, `Clfinal' as described in the text,  obtained from the sky region survived after application of mask M6 and after bias correction, 
is shown in points with brown color along with the error-bars. Theoretical power spectrum as obtained from Planck-2015 analysis
is plotted in blue line for comparison.}
\label{cl_plot}
\end{figure*}

\begin{figure*}
\includegraphics[scale=0.6]{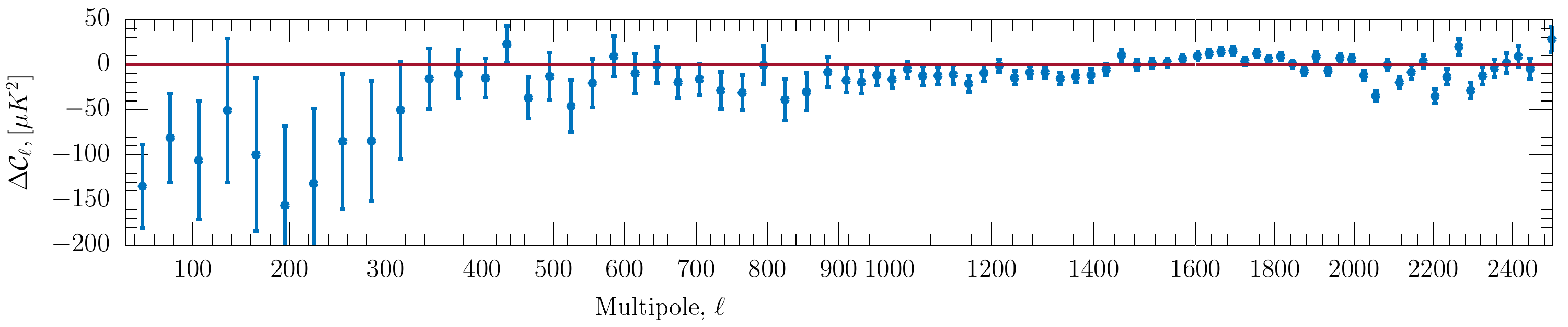}
\caption{Difference of our bias corrected binned power spectrum (shown in Fig.~\ref{cl_plot}) from the Planck-2015 
binned angular power spectrum for the multipole range $47 \le \ell \le 2500$ is shown  with the error bars computed from 
Monte-Carlo simulations.}
\label{cl_diff}
\end{figure*}

\begin{figure}
\includegraphics[scale=0.7]{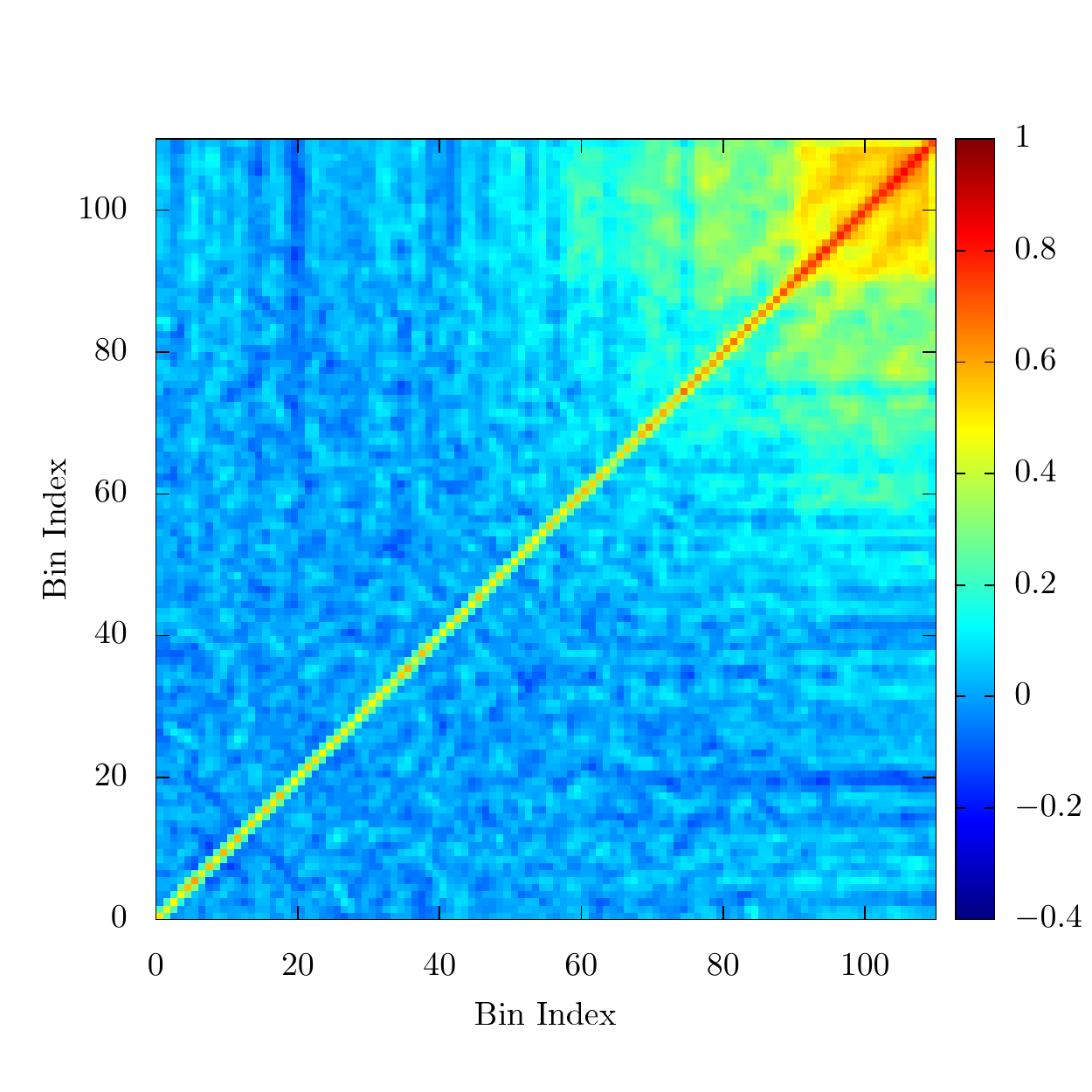}
\caption{Correlation of CMB angular power spectrum, `Clfinal',  shown in Fig~\ref{cl_plot} between different multipoles and bins.}
\label{corr_cl}
\end{figure}

We generate CMB angular power spectrum  using cross-correlation of cleaned maps ${\tt C_1P_2}$ and ${\tt C_2P_2}$
using MASTER approach as described in~\cite{Hivon2002}. We define a `M6' mask (shown in Figure~\ref{M6Mask}) that excludes all regions near the galactic 
plane defined by region 1 to 6 of Figure~\ref{intensity_mask} along with the position of known point sources. We obtain power 
spectrum, $C^m_{\ell}$, of this mask and use it to obtain the mode-mode coupling matrix, $M_{\ell \ell'}$ following, 
\begin{eqnarray}
M_{\ell \ell'} = \frac{2\ell'+1}{4\pi}\sum_{\ell''}(2\ell''+1)C^m_{\ell''}
  \begin{pmatrix}
    \ell  & \ell' & \ell'' \\
    0     &   0   &   0
  \end{pmatrix}^2
\end{eqnarray}
where the last term on the right had, side denotes the Wigner-3j symbol.  We apply
M6 mask on ${\tt C_1P_2}$ and ${\tt C_2P_2}$ maps and obtain the partial cross power spectrum, $\hat {\tilde C}_{\ell}$. We
convert the partial sky cross-power spectrum to the full sky estimates of CMB angular power 
spectrum, $\hat C_{\ell}$ by first inverting the mode-mode coupling matrix and then computing, 
\begin{eqnarray}
\hat C_{\ell} = \frac{ M^{-1}_{\ell, \ell'} \hat {\tilde C}_{\ell}}{P^2_{\ell}B^1_{\ell}B^2_{\ell}}
\label{Cl0}
\end{eqnarray}     
where $P_{\ell}$ denotes the pixel window function and $B^1_{\ell}$ and $B^2_{\ell}$ are respectively 
beam window function of the 353 GHz maps corresponding to sets $S_1$ and $S_2$. We denote this power spectrum 
by `Cl0'. 

As described in Section~\ref{simulations} (also see ~\cite{Saha2008}) the ILC power spectrum contains 
a negative bias which is strongest at the lowest multipoles. We estimate the negative bias in our power 
spectrum  estimated from the sky region, survived after masking by `M6' mask, using Monte-Carlo simulations. 
We correct for this negative bias in `Cl0' cross-spectra for the multipole range $2 \le \ell < 500$. For the 
multipole range $ 500 \le \ell \le 3000$ `Cl0'  cross-power spectrum contains a residual unresolved 
point source contamination. We first estimate a model for the residual unresolved point source contamination 
in the `Cl0' spectrum by fitting the excess with respect to a fiducial CMB power spectrum~\citep{PlanckCosmoParam2016} 
with the simple phenomenological model, 
\begin{eqnarray}
C^e_\ell = C\left(\frac{ \ell}{2000}\right)^\alpha +  C'\left(\frac{ \ell}{2000}\right)^\gamma
\label{highl_excess}
\end{eqnarray}     
where $C^e_\ell$ denotes the excess power at multipole $\ell$, $C, C', \alpha, \gamma$ are constants which 
are obtained by minimizing the $\chi^2$ statistic defined by, 
\begin{eqnarray}
\chi^2 = \sum_{\ell=500}^{\ell=3000}\left(\frac{\hat C_\ell-C_\ell^{fid}}{\sigma_\ell} \right)^2
\end{eqnarray}
where $\hat C_\ell$ is given by Eqn~\ref{Cl0} and $\sigma_\ell$ represents the  standard deviation of power spectrum 
 `Cl0' which we estimate from the diagonal elements of the multipole space covariance matrix. From the fit we obtain
$C = 84.082 \pm 12.82, \alpha = 1.7793 \pm 0.4659, C' = 44.9187 \pm  11.48$ and  $\gamma = 7.83475 \pm 0.5524$, 
with a value of reduced $\chi^2 = 5.77$. We use these best-fit parameter values in Eqn~\ref{highl_excess}  and correct 
`Cl0' cross power spectrum  by  subtracting the excess as determined by this equation for the multipole range 
$500 \le \ell \le 3000$.   Using the bias corrected `Cl0' power spectrum
we produce our final CMB angular power spectrum, `Clfinal',  as follows.  For $2\le \ell \le 29$ we take power spectrum at each 
multipole $\ell$. For $2500 \ge \ell > 29$ we perform  a binning identical to Planck-2015 results. The resulting 
power spectrum is shown in Fig~\ref{cl_plot} in brown colored points, along with the Planck-2015 binned power spectrum 
(shown in green points). We have also shown in figure Planck-2015 best fit theoretical CMB power spectrum~\citep{PlanckCosmoParam2016} 
for comparison. In Fig~\ref{cl_diff} we have shown difference of `Clfinal' and Planck-2015 binned power spectrum
for the bin-middle values starting from $\ell = 47$, along with the error-bars as computed from the Monte-Carlo
simulations. We see from this figure, `Clfinal' contains less power for mutipoles 
$47\le \ell \le 300$. 

We  estimate the full multipole space covariance matrix of the bias corrected `Cl0' angular  power spectrum using 
Monte-Carlo simulations. The error-bar on unbinned and binned power spectrum at each multipole $\ell$ 
and at each bin is given by the  square root of the diagonal elements of such covariance matrices. We show the  
correlation matrix, of the power spectrum shown in Fig~\ref{cl_plot} in Fig~\ref{corr_cl}. Between $0$ to $28$ 
bin indices of this figure represent multipoles $\ell =2 $ to $\ell = 29$. For bin indices between $29$ to 
$110$ correlation between binned spectra for different Planck bins are represented.    
 
\section{Simulations}
\label{simulations}
We validate the methodology of the new iterative ILC algorithm by performing Monte-Carlo simulations 
of the entire foreground removal and power spectrum estimation pipeline. Using the results from the 
simulations we understand foreground leakage, ILC negative bias~\citep{Saha2008} and positive bias due to residual 
foregrounds in the final CMB cross power spectrum.  Further, we estimate covariance matrix of final CMB 
power spectrum in the multipole space using the simulations. We use the diagonal elements of the 
covariance matrix to estimate the error bar on our final power spectrum. Below, we first describe different 
components used in the simulations. We then discuss the results of our simulations.    

\subsection{CMB Component}
We generate $160$  random realizations of CMB temperature anisotropy maps compatible to Planck LCDM 
power spectrum~\citep{PlanckCosmoParam2016} at $N_{\textrm side} = 2048$. We downgrade the pixel resolution 
of these parent set of CMB maps to the corresponding pixel resolution of each map belonging to set $S_1$ and 
$S_2$ as mentioned in Table~\ref{ListMaps}. Each of these maps is then smoothed by the instrumental beam function (e.g., see Fig~\ref{Bl})  of the 
corresponding detector (or  detector set) map. 
 
\subsection{Foreground Components}
The galactic foreground model used in our simulations consists of three major foreground components, namely, 
synchrotron, freefree and thermal dust. We use Planck-2015 foreground model~\citep{Planck2015_fg}  and their published 
templates (provided by the Planck science team at $N_{\textrm side} = 256$ and beam resolution $1^\circ$) to 
estimate the emission levels of these foreground components at different Planck and WMAP frequencies. 

One of the major concern for reliable removal of foreground components is variation of spectral indices 
of synchrotron and thermal dust components over the sky. Although,~\cite{Planck2015_fg} provided a spectral 
index map for the thermal dust component we do not have any template map for spectral index of synchrotron 
component from joint analysis of WMAP and Planck observations. To generate the synchrotron spectral index 
map at various WMAP and Planck frequencies we therefore first create a  synchrotron spectral index map 
using WMAP K1 and Planck 30 GHz frequency maps.  

\subsubsection{Synchrotron Spectral Index Map} 
\label{SpIndexMap}
\begin{figure}
\centering
\includegraphics[scale=0.33]{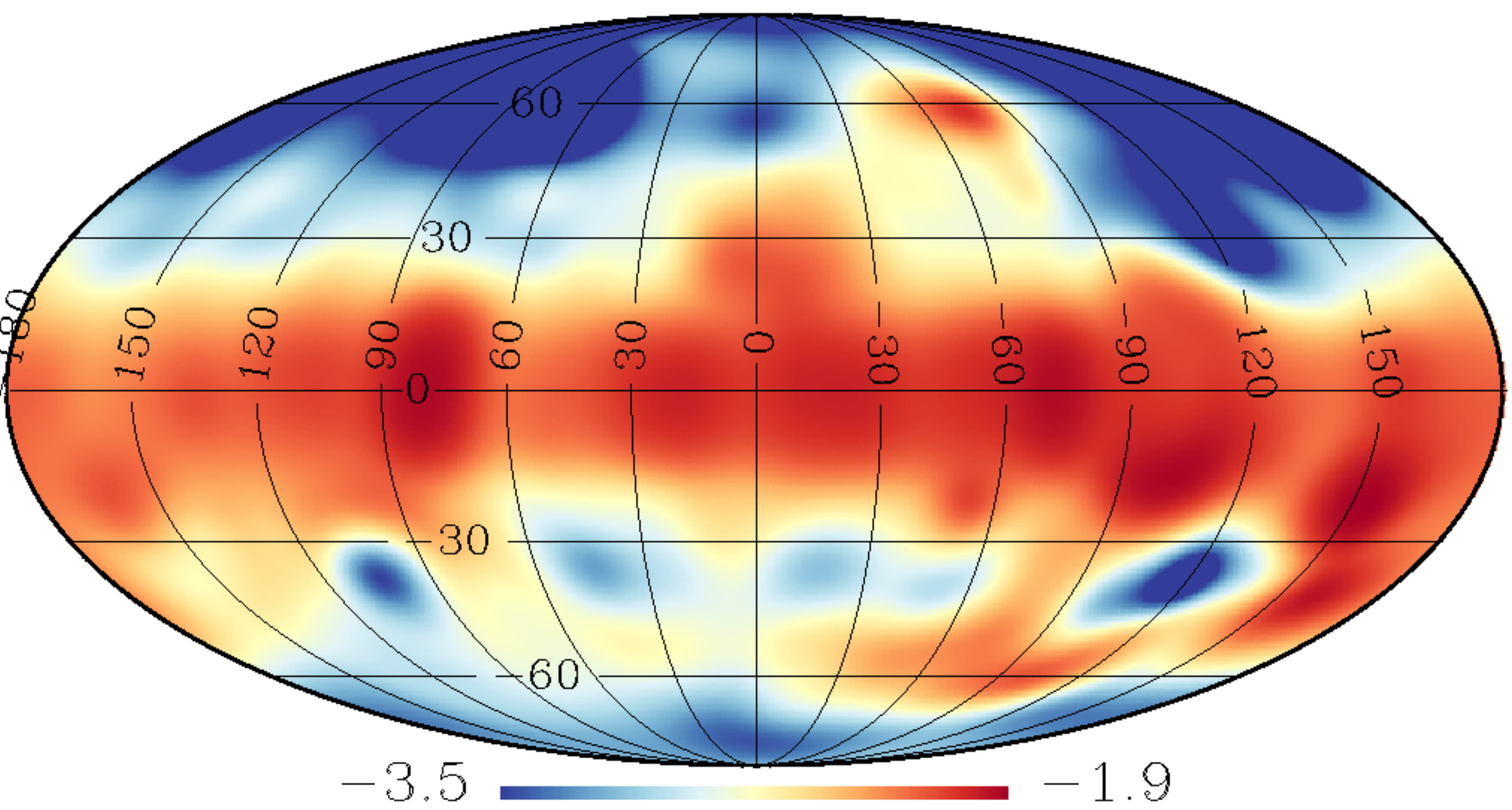}
\caption{Synchrotron spectral index map, $\beta_{s256}(p)$ at $N_{\textrm{side}}$ = 256 used to estimate synchrotron emission 
levels at various WMAP and Planck frequencies of this work.}
\label{beta_synch}
\end{figure}

To form the  synchrotron spectral index  map we take  WMAP 23 GHz (K1 band, $N_{\textrm{side}}=512$)  
and Planck 30 GHz ($N_{\textrm {side}} = 1024$) frequency maps, convert their pixel temperature values in $\mu K$ 
(thermodynamic) unit, and downgrade their pixel resolution to $N_{\textrm{side}} = 256$. We smooth these low 
pixel resolution maps by a Gaussian beam function of FWHM = $1^\circ$ by multiplying their spherical harmonic 
 coefficients by the ratio of beam window functions of $1^\circ$ Gaussian window to the window function of 
each map's native resolution beam function.  We  also downgrade pixel resolution of the Planck {\tt COMMANDER}
CMB map to $N_{\textrm{side}}= 256$. Since the {\tt COMMANDER} CMB   map is provided in a Gaussian beam resolution 
of $5^\prime$ we smooth the low pixel resolution CMB map to $1^\circ$ beam resolution by multiplying its 
spherical harmonic coefficients  by the beam window function of a Gaussian beam of FWHM = $\sqrt{60^2-5^2} = 59.79^\prime$ 
and convert it to $\mu K$ temperature unit. We subtract from the $23$ and $30$ GHz frequency maps  at $N_{\textrm{side}} = 256$
the {\tt COMMANDER} CMB map, Planck best fit thermal dust and freefree maps derived at these  two frequencies
\footnote{A detailed discussions on how to generate the thermal dust and freefree emissions at different Planck and WMAP 
frequencies are mentioned later in this Section.}. Ignoring spinning dust emission 
the two resulting maps at $23$ and $30$ GHz contain only synchrotron component. We further smooth these two maps by the beam 
function of a Gaussian window of FWHM = $\sqrt{900^2-60^2}$  to bring them to a resolution of $15^\circ$. 
We denote the resulting low resolution maps at $23$ and $30$ GHz respectively as $M_{23}$ and $M_{30}$. We form the  
synchrotron spectral index map, $\beta_{s}(p)$,  following,
\begin{eqnarray}
 \beta_{s}(p) = -\ln\frac{a_{30}M_{23}}{a_{23}M_{30}} \times \ln\frac{30}{23}
\end{eqnarray}  
where $a_{23}$ and $a_{30}$ are respectively antenna to thermodynamic temperature conversion factors 
at the indicated frequencies. Since the natural logarithm of zero or a negative number is undefined, in principle
one should restrict to only those pixels for which both maps $M_{23}$ and $M_{30}$ are positive definite. 
We, however, note that, due to relatively larger smoothing window present in these maps they are already 
synchrotron signal dominated and hence we find no zero or negative pixels in these maps. The maximum and  
minimum values of the spectral index map are respectively $-1.933$ and $-11.475$. Our final synchrotron 
spectral index map, $\beta_{s256}(p)$, at $N_{side} = 256$ is obtained by replacing pixels of this map  
less than $-3.50$ by  $-3.50$. We have shown the synchrotron spectral index map in Fig~\ref{beta_synch}.     
 
To generate the synchrotron emission map, $S_{\nu}(p)$ at different WMAP and Planck frequencies at $N_{\textrm {side}} = 256$ we 
first extrapolate the Planck synchrotron template (provided at $\nu = 408$ MHz, in $\mu K$ antenna temperature unit) 
at 23 GHz using a constant  spectral index, $\beta_s = -2.80$, all over the sky.  We call this template $S_{23}(p)$. This 
new template is now extrapolated at all WMAP and Planck frequencies following,
\begin{eqnarray}
S_{\nu}(p) =S_{23}(p)\left(\frac{\nu}{23.0}\right)^{\beta_{s256}(p)}
\end{eqnarray}
where $\beta_{s256}$ represents the spectral index map described above. Thus our 
synchrotron emission maps consist of a variable spectral index model  at WMAP and Planck frequencies. 

We generate freefree signal at different frequencies using its  model prescribed in the third column of Table 4 of 
~\citep{Planck2015_fg}. We note that both {\it emission measure} and electron temperature $T_e$ are functions 
of locations on the sky in this model. For thermal dust component we use the variable spectral index and 
dust temperature model as given in the  Table 4 of~\citep{Planck2015_fg}.    

All the three components for different frequencies are now smoothed by the beam window functions of 
different detector and detector sets of WMAP and Planck observations. We upgrade the pixel resolution of 
the smoothed maps to the pixel resolutions as mentioned in the last column of Table~\ref{ListMaps}.  
We add the three components at these new pixel resolutions and convert the resulting {\it net} foreground 
emission maps at different frequencies to thermodynamic ($\mu K$) temperature unit.

\subsection{Detector Noise} 
To generate the noise maps at WMAP frequencies we use the noise per observation ($\sigma_0$) values for Stokes I  
observations  as given Table 5 of~\cite{Hinshaw2013} along with the total number of observations for 
a pixel $p$, $N_{\textrm obs}(p)$, as provided by the WMAP science team with the individual detector set maps\footnote{Available from 
https://lambda.gsfc.nasa.gov/product/map/dr5/maps\_da\_r9\_i\_9yr\_get.cfm.}. The noise at pixel $p$ at a frequency 
map $\nu_i$ is given by, 
\begin{eqnarray}
n^i_(p) = \frac{\sigma^i_0}{\sqrt{N^i_{\textrm obs}}} G
\end{eqnarray} 
where $G$ denotes a  Gaussian deviate with zero mean and unit variance and $\sigma^i_0$ represents the 
noise level per observation for the differencing assembly with frequency  $\nu_i$ . For both LFI and HFI Planck maps we create  
noise realization by using the intensity noise variance values given in the fifth columns of the respective frequency
band map files.  We create a total of $160$ noise realizations for each set $S_1$ and $S_2$.
The noise properties of different detector and detector set maps are uncorrelated with each other for all realizations. 
For WMAP DA maps following the order from  K1 to W4,  mean angular power spectrum estimated from these $160$ 
noise maps and and averaged over  all multipoles are respectively $7.181$E-3, $7.254$E-3, $13.887$E-3, $12.495$E-3, $21.726$E-3, 
$17.206$E-3, $45.381$E-3, $56.398$E-3, $63.109$E-3 and $59.686$E-3 $\mu K^2$.  For Planck 30 GHz, 70 GHz, 100 GHz, 143 GHz, 217GHz and 
353 GHz corresponding  mean  angular power spectrum  of the noise maps are respectively, $3.220$E-3, $3.359$E-3, $5.094$E-4, $9.410$E-5,
$1.865$E-4 and $2.0153$E-3 $\mu K^2$. Thus Planck 143 GHz has the least noisy observations. Among the two highest frequency 
maps $217$ GHz has less noise level than the $353$ GHz maps.  
\subsection{Simulated Frequency Maps}

\begin{figure}[t]
\centering
\includegraphics[scale=0.35]{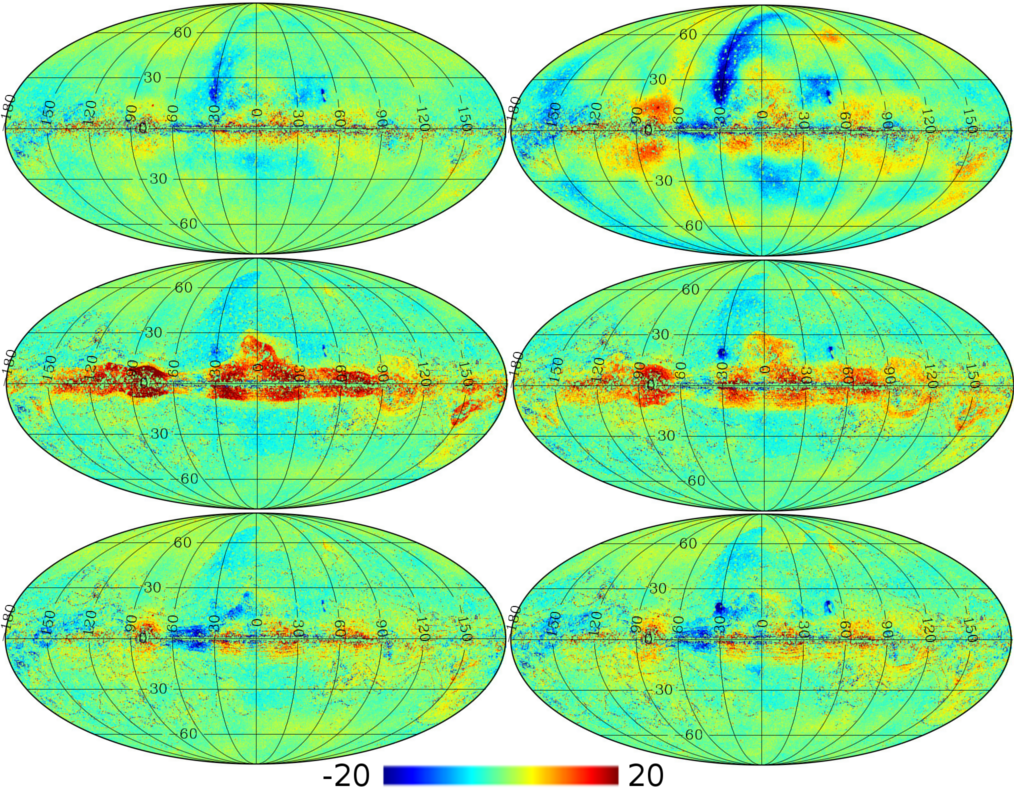}
\caption{Figure showing the foreground residuals for phase 1 (top), phase 3 (middle) 
and phase 2 (bottom) analysis in set $S_1$ (left) and $S_2$ (right) cleaned maps. The plots of the  
middle panel dominated by the foreground leakage signal. No leakage is present in the figures 
of bottom panel. }
\label{ResMap_ns2048}
\end{figure}

\begin{figure*}[t]
\includegraphics[scale=0.65]{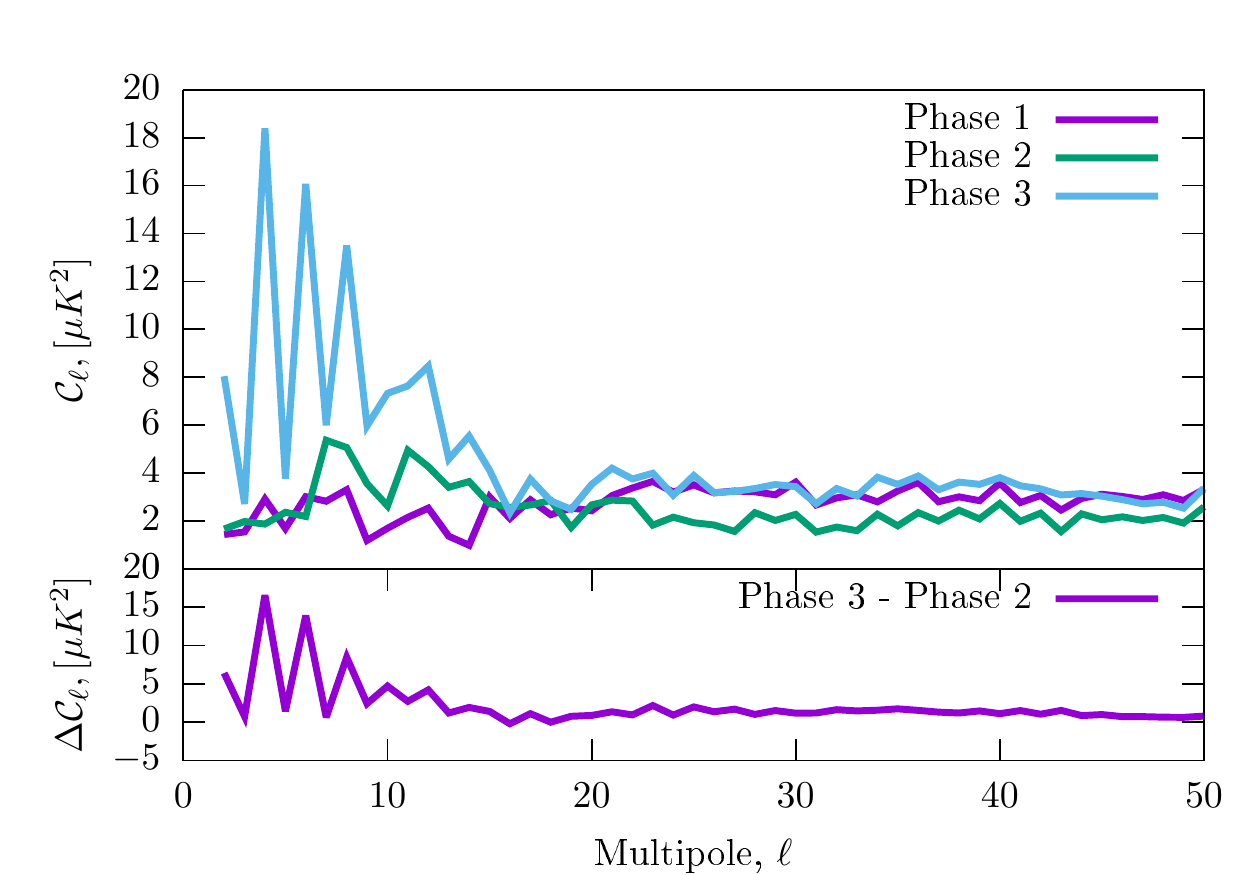}
\includegraphics[scale=0.65]{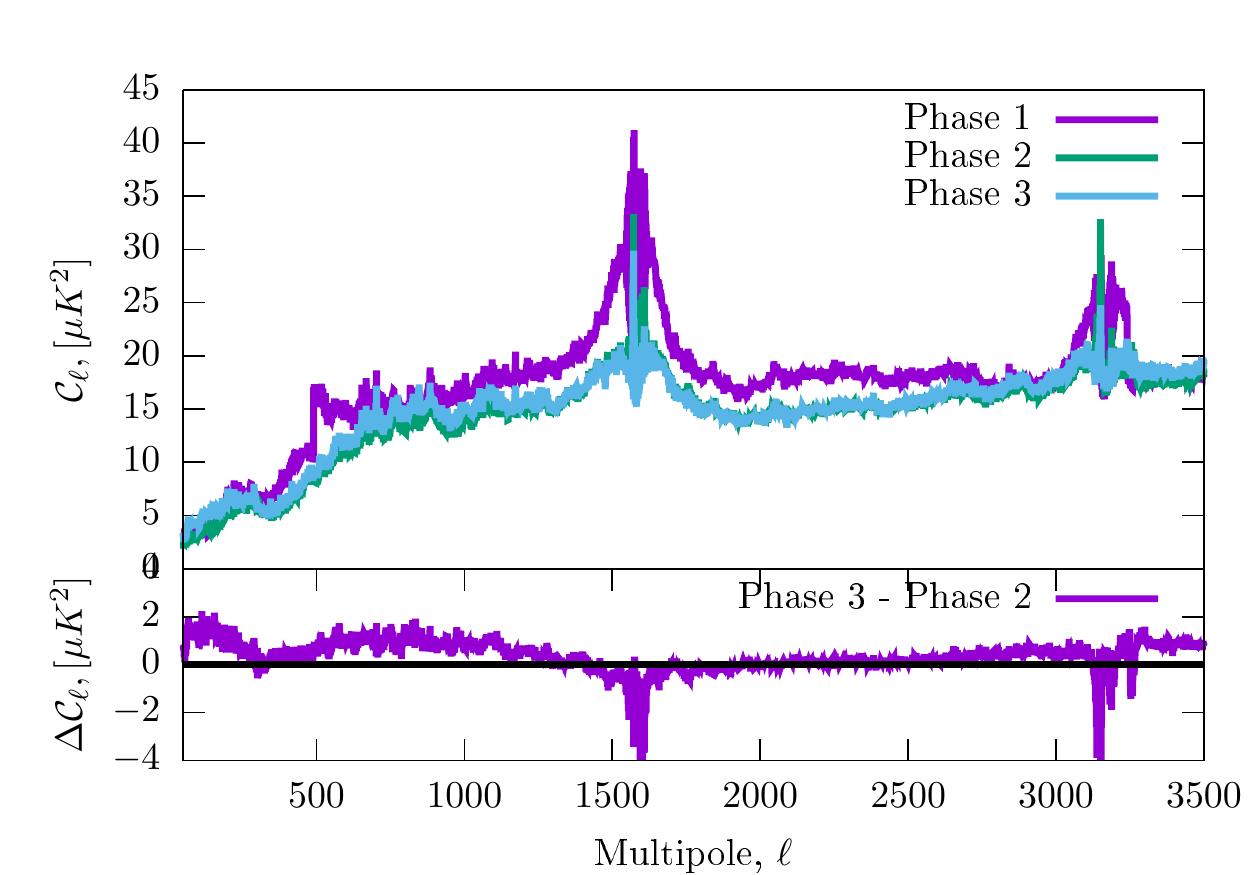}
\caption{{\it Left:} Cross power spectrum $\mathcal C_{\ell} = \frac{{\ell}(\ell +1)}{2\pi}C_{\ell}$, of foreground leakage signal from  pairs 
of cleaned maps obtained from different phases of analysis as described in the text, for the multipole range $2 \le \ell \le 50$. The lower 
part shows the  difference in power spectrum between the phase 3 and phase 2 analysis, indicating foreground leakage in phase 2, 
but no foreground leakage in phase 2 analysis. 
{\it Right:} Same as left panel, but for higher multipole range, $50 \le \ell \le 3500$.} 
\label{leak_power_ns2048}
\end{figure*}

Once CMB, foreground and detector noise realizations have been created we simply add them to form $160$ realizations of WMAP and 
Planck detector and detector set maps at the native pixel resolution of different maps corresponding to sets $S_1$ and $S_2$. 
These realizations, henceforth labeled as, set `Sim0', represent the simulations of signal (in thermodynamic $\mu K$ temperature unit) that is 
actually measured by the WMAP and Planck satellite missions.   

Since we perform our analysis after converting all WMAP and Planck frequency maps to  $N_{\textrm side} = 2048$, 
 we form another set of frequency maps for all $160$ realizations, labeled as `Sim2048' by transforming the 
`Sim0' maps to spherical harmonic space and then using Eqn~\ref{general_ilc} upto $\ell^{(cut)i}_{max}$ values 
as listed in Table~\ref{ListMaps}.  This brings all the maps of set `Sim2048' to the same beam and pixel 
resolution upto the chosen $\ell^{(cut)i}_{max}$ values. Since point sources dominate on small angular scale on CMB 
maps, we further mask the locations of resolved point sources from the maps of the set `Sim2048' using `PSMask'. 
This is justified since we are interested to remove diffuse foregrounds. Further, masking out the positions of 
resolved point sources guarantees that the weights are not affected by these sources, which leads to better efficiency in 
diffuse foreground removal. We label the set of point source masked realizations as `Sim2048m'.

\subsection{Results and Analysis}
Using the input maps of set `Sim2048m' we perform $160$ Monte-Carlo simulations of entire pipeline of foreground 
removal and power spectrum estimation technique, for both phase 1 and phase 2 analysis, following the same procedure as followed for the case 
of real data. We also perform $160$ Monte-Carlo simulations of the old iterative ILC foreground removal and power 
spectrum estimation as in phase 3 analysis of the data.  Since the input frequency maps for all simulations contain 
varying spectral indices for both synchrotron and thermal dust components, moreover, each frequency map contains 
detector noise contamination, foreground removal in ILC method cannot be completely efficient~\citep{Saha2008}. 
This leads to residual contamination. Further, the old iterative ILC 
method suffers from the problem of foreground leakage as discussed in Section~\ref{leakage}.   

To understand the residuals present in the cleaned CMB maps obtained at the end of  different phases of analysis, 
we form  difference maps by subtracting the input CMB maps from cleaned maps obtained from simulation. We make an average 
difference map using all such difference maps for each set $S_1$ and $S_2$ and for each phase using a total of $160$ Monte-Carlo simulations. Since foreground 
removal in phase 1 involves only one iteration over the entire point source masked sky, and spectral indices 
of both synchrotron and thermal dust vary with sky positions, there exist some foreground  residuals due to 
imperfect removal foregrounds in the averaged cleaned map $\tt C_1P_1$ and $\tt C_2P_1$. In the top row of Fig~\ref{ResMap_ns2048} we show 
the residual foreground present in the cleaned map of phase 1 analysis. Top left figure of the top panel shows 
residual for the cleaned map $\tt C_1P_1$ and top right figure shows the residual for the cleaned map 
$\tt C_2P_1$. In  both the maps of top panel we see the residuals present along the galactic plane and along 
the north polar spur. We note that since in  phase 1 analysis the entire point source masked region was cleaned in a 
single iteration foreground leakage is absent in these plots.  

In the phase 3 foreground removal, we have simply followed the old ILC algorithm in an iterative fashion 
in the multipole space. The cleaned maps from different Monte-Carlo simulations now contain foreground 
residuals that arise due to imperfect foreground cleaning due to variation of  foreground spectral indices
within each region plus the foreground leakage as described in Section~\ref{leakage}. Contributions due to 
both these type of residuals are manifested in the galactic region. Such residuals are shown in the middle 
panel of the Fig~\ref{ResMap_ns2048} for phase 3, set $S_1$ (left) and phase 2 set $S_2$ cleaned map (right)
respectively.   

Following the discussion of Section~\ref{leakage} in the phase 2 cleaned maps the foreground leakage 
completely stops. The foreground residuals in the phase 2 cleaned maps then arise due to imperfect 
foreground cleaning due to spectral index variation within any given region. Plots of such residuals 
are shown in the average difference of output and input CMB maps corresponding to sets $S_1$ and $S_2$ 
computed from our Monte-Carlo simulations at the bottom panel of Fig.~\ref{ResMap_ns2048}. Comparing the 
middle and bottom panel of this  figure we see improvements achieved by extending the usual iterative 
ILC approach in the multipole space to the two-phase  approach as proposed in this work. A difference 
of middle and bottom panel maps in Fig which cleanly shows the foreground leakage signal.    

We quantify the foreground residuals that originate due to imperfect foreground removal as well as foreground leakage, 
by estimating angular power spectrum of the difference maps which are shown in Fig~\ref{ResMap_ns2048}.
We show the power spectrum of set $S_1$ residual maps for all three phases in Fig.~\ref{leak_power_ns2048}. 
As shown from the top left and top right panels of this figure the phase 1 (set $S_1$) cleaned map contains 
more residuals, in general for multipole $\ell \ge 400$ compared to both phase 2 and phase 3 cleaned maps.
This clearly demonstrates advantage of performing foreground removal in multiple iterations over a single 
iteration. At the large scale on the sky, for multipoles, $7\le \ell \le 14$, phase 1 (set $S_1$) cleaned 
map contains less foreground residuals than both phase 2 and phase 3 cleaned maps.  Comparing results from 
the phase 2 and phase 3 analysis we see that our new method 
performs  better than the old iterative ILC method in multipole space for almost all multipoles. For $\ell \le 20$ (e.g.
see top and bottom panels at left of Fig.~\ref{leak_power_ns2048}) the new method performs significantly better than 
the old iterative ILC method. Moreover, the phase 2  results have less foreground residuals over single-iteration 
phase 1 analysis for multipoles, $\ell > 20$.  

\begin{figure*}[t]
\includegraphics[scale=0.7]{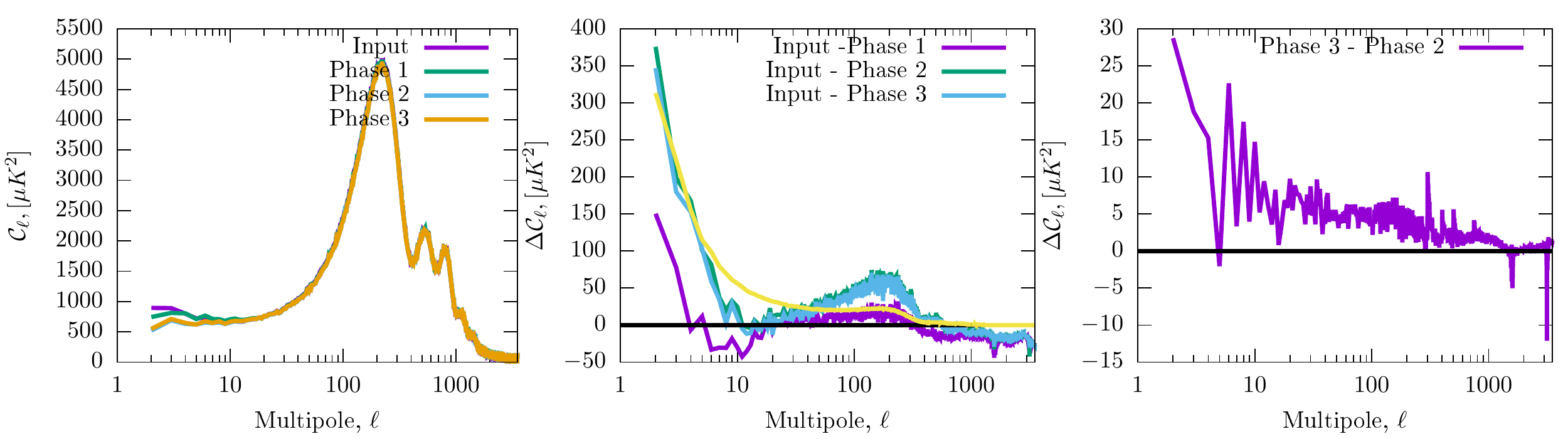}
\caption{{\it Left:} Figure showing the average cleaned cross power spectra  for different 
phases of analysis of this work, along with the average CMB cross power spectrum, computed 
from sky region survived after applying `PSMask', using $160$ Monte-Carlo simulations. 
All the spectra of this plot have beam and pixel effects present. {\it Middle:} 
Figure showing the magnitude of the negative bias in  cross power spectra of different phases, plotted 
in the left figure. {\it Right:} Difference between the phase 3 and phase 2 cross power 
spectrum. In all the three plots, $\mathcal C_{\ell} = \frac{{\ell}(\ell +1)}{2\pi}C_{\ell}$ 
is plotted along the power spectrum axes.}
\label{cl_bias}
\end{figure*}

What is the total bias in the ILC  cross power spectra obtained from the analysis of the individual three 
phases? The ILC power spectrum contains a negative bias which is most dominant at the lowest multipoles as 
discussed in~\cite{Saha2008}. Since the bias due to foreground residuals becomes positive one has to be 
cautious when interpreting net bias in ILC power spectrum. A net positive bias may not indicate presence 
of only residual foreground bias, similarly a net negative bias may not merely indicate the actual magnitude
of the negative bias at any given $\ell$.  Since we do not know a-priori at which multipole a foreground bias 
will be present, neither do we know  exact analytical form of negative ILC bias for the iterative method of ILC
foreground removal\footnote{In~\cite{Saha2008} 
the authors provided an expression of the negative bias only for a full sky single-iteration analysis.}, it becomes difficult to 
estimate magnitude of individual positive and negative biases at any multipole $\ell$. However, one can find from a plot of net bias 
different multipole ranges where individual positive and negative biases dominate.  In left panel Fig.~\ref{cl_bias} we 
have shown the average cleaned cross power spectra of the two sets at any given phase, for all the three 
phases of analysis of this paper  compared against the average CMB input power spectrum. All the spectrum of this 
plot have both pixel and beam effect present.  All  four cross spectra  of this plot are estimated from the portion 
of the sky after masking by the `PSMask'. In the middle panel of this figure we have shown the negative of net bias in the 
phase 1, 2 and 3 cross spectra which are plotted in the left panel. At the multipole range $\ell \le 15$ we see that the bias 
in phase 2 and 3 cross-spectra are more negative than the phase 1 cross power spectra. This indicates that the 
phase 1  cross-power spectra in this multipole range contains a foreground residual making the net bias in it less negative. 
In fact it it is likely that the foreground residual exists in phase 1 cross-spectra for all multipoles upto $\ell \le 300$. 
Since the negative bias is correlated with the CMB angular power spectrum there exists some negative bias at the first 
acoustic peak  for all, phase 1, phase 2 and phase 3 cross-power spectra. In the middle panel of this figure we have also
plotted $\frac{C^{fid}_{\ell}\ell(\ell+1)}{(2\ell+1)2\pi}$, where $C^{fid}_{\ell}$ represents a fiducial CMB power 
spectrum~\citep{PlanckCosmoParam2016}, in yellow color as an intuitive understanding of variation of 
negative bias with multipoles that we expect from a full sky foreground removal involving only one foreground component
with constant spectral index all over the sky and in an ideal experiment where no detector noise is present.  In the 
right most panel of this figure we show the difference of bias for phase 2 and phase 3 cross power spectrum. Assuming the negative bias 
is comparable for both phase 2 and phase 3 cross-power spectra we conclude that the phase 2 analysis contains significantly less 
foreground residual power for multipole range $\ell \le 1000$. This clearly demonstrates that the new ILC  technique 
proposed in this  work improves the old iterative technique.

\section{Discussion and Conclusion}
\label{D&C}
In the usual practice an ILC foreground minimization is performed in the multipole space to take into account 
variation of foreground spectral properties (or variation of detector noise level) with  multipoles. Foreground 
spectral properties however not only depend upon the multipoles but also on the sky regions under consideration, 
since the populations foreground emitting sources varies in their intrinsic characteristics with their locations 
in the Milky Way galaxy and hence with the directions on the sky. One therefore choses to perform the usual 
ILC technique in multipole space successively over different regions of the sky, which are so created that the 
nature of foreground properties remains approximately similar over any given region. This so-called multi-scale 
and multi-region approach of sky however suffers from a major limitation of leakage of foreground signal from the 
uncleaned region to the region that is currently being cleaned at any given iteration. {\it In this work, we have 
for the first time, discovered the leakage signal by formulating an analytical framework.}. We have extended the 
usual ILC approach into a new two-phase analysis so that the leakage  signal completely stops in the new method.    

We have applied the new approach on the Planck-2015 data sets  and WMAP 9-year frequency maps to estimate a pair of diffuse 
foreground minimized CMB temperature anisotropy map with the independent detector noise properties from the 
regions of the sky left after masking the positions of the known point sources. {\it Both the cleaned maps obtained 
by us have higher resolution than the Planck-2015 foreground cleaned CMB maps.} We estimate angular cross power spectrum 
of these two foreground minimized maps to estimate the CMB angular power spectrum from  the region of the sky defined by 
so-called `M6' mask. Our final power spectrum, `Clfinal',  matches well with the Planck-2015 published power spectrum~\citep{PlanckCosmoParam2016}.  
with some differences at different regions of multipoles. 

We have performed a detailed Monte-Carlo analysis with Planck foreground templates for synchrotron, free free and thermal dust 
components along with the detector noise levels compatible to Planck and WMAP observations. The foreground models used 
in the simulations takes into account complex scenario such as spectral index variations of synchrotron and thermal dust components 
with the angular positions on the sky.  The Monte-Carlo analysis 
confirms the foreground leakage signal for the case of old iterative ILC algorithm, in the multipole space, and shows 
that  the new multi-phase iterative ILC algorithm gets rid of foreground leakage signal as proved in Section~\ref{leakage}.
We use the Monte-Carlo simulations to estimate the biases in different phases of ILC algorithm, and shows that the 
the new method has the less net bias (due to ILC negative bias and the positive foreground bias) than the old, at all 
angular scales, before the biases for both methods become negligible at high $\ell$.  We use the Monte-Carlo simulations 
to estimate the error-bar of the final power spectrum estimated from the Planck and WMAP data.

{\it Removing the foreground leakage signal in the new multi-phase algorithm will enable us to perform even more reliable analysis 
of CMB in terms of its power spectrum. The improvements of this work is expected to bear a very important implication 
on CMB polarization analysis, where the effect of the leakage is expected to be more significant, since the polarized 
diffuse emissions, specifically near the galactic plane, may be way stronger than the weak primordial CMB signal.}  

The new ILC algorithm as implemented  in this paper has some additional advantages also. It can now handle different 
input frequency maps with equal footing by first converting them to the resolution function of the highest resolution 
input frequency map  upto certain maximum multipoles $\ell^{(cut)i}_{max}$ which are decided by the resolution of each  input 
frequency maps and simultaneously upgrading them to the highest pixel resolution of all available frequency maps.  
This avoids any need to deconvolve partial sky cross-power spectra covariance matrices by the beam window function 
during the weight estimation stage of ILC algorithm. Since division of  the partial sky cross covariance matrix 
by the beam  functions for deconvolution purpose is not a mathematically well-defined  operation, we consider avoiding such operations is 
an important improvement to the ILC algorithm.  We bring the beam and pixel resolution of all input frequency maps at the 
highest resolution at the very beginning of the analysis. This allows to perform the foreground removal  
on the part of the sky, after masking off any undesired region for foreground cleaning, which may contain so strong 
foreground emissions that they cannot be cleaned effectively for the purpose of estimation of  cosmological signal. {\it This helps 
to remove foregrounds from the rest of the unmasked sky portions with better efficiencies since the weights are now 
not affected by the strongly contaminated regions.}

R.S. acknowledges financial support under the research grant SR/FTP/PS-058/2012 by SERB, DST. P.K.A. is supported 
by Claude Leon Post-doctoral Fellowship program of Claude Leon Foundation, South Africa 
at UCT. We use publicly available HEALPix~\cite{Gorski2005} package available from 
http://healpix.sourceforge.net to perform forward and backward spherical harmonic transformations and 
for visualization purposes. We acknowledge the use of the Legacy Archive for Microwave Background 
Data Analysis (LAMBDA). LAMBDA is a part of the High Energy Astrophysics Science Archive Center (HEASARC). 
HEASARC/LAMBDA is supported by the Astrophysics Science Division at the NASA Goddard Space Flight
Center.


\begin{thebibliography}{}
\expandafter\ifx\csname natexlab\endcsname\relax\def\natexlab#1{#1}\fi

\bibitem[{{Bennett} {et~al.}(2003){Bennett}, {Hill}, {Hinshaw}, {Nolta},
  {Odegard}, {Page}, {Spergel}, {Weiland}, {Wright}, {Halpern}, {Jarosik},
  {Kogut}, {Limon}, {Meyer}, {Tucker}, \& {Wollack}}]{Bennett03_fg}
{Bennett}, C.~L., {Hill}, R.~S., {Hinshaw}, G., {et~al.} 2003, Astrophys. J.
  Suppl. Series, 148, 97

\bibitem[{{Bennett} {et~al.}(2013){Bennett}, {Larson}, {Weiland}, {Jarosik},
  {Hinshaw}, {Odegard}, {Smith}, {Hill}, {Gold}, {Halpern}, {Komatsu}, {Nolta},
  {Page}, {Spergel}, {Wollack}, {Dunkley}, {Kogut}, {Limon}, {Meyer}, {Tucker},
  \& {Wright}}]{Bennett2013}
{Bennett}, C.~L., {Larson}, D., {Weiland}, J.~L., {et~al.} 2013, \apjs, 208, 20

\bibitem[{{Bielewicz} {et~al.}(2012){Bielewicz}, {Banday}, \&
  {G{\'o}rski}}]{Bielewicz2012}
{Bielewicz}, P., {Banday}, A.~J., \& {G{\'o}rski}, K.~M. 2012, \mnras, 421,
  1064

\bibitem[{{Bouchet} \& {Gispert}(1999)}]{Francois1999}
{Bouchet}, F.~R., \& {Gispert}, R. 1999, \na, 4, 443

\bibitem[{{Bouchet} {et~al.}(1999){Bouchet}, {Prunet}, \&
  {Sethi}}]{Francois1999a}
{Bouchet}, F.~R., {Prunet}, S., \& {Sethi}, S.~K. 1999, \mnras, 302, 663

\bibitem[{{Bunn} {et~al.}(1994){Bunn}, {Fisher}, {Hoffman}, {Lahav}, {Silk}, \&
  {Zaroubi}}]{Bunn1994}
{Bunn}, E.~F., {Fisher}, K.~B., {Hoffman}, Y., {et~al.} 1994, \apjl, 432, L75

\bibitem[{{Calabrese} {et~al.}(2013){Calabrese}, {Hlozek}, {Battaglia},
  {Battistelli}, {Bond}, {Chluba}, {Crichton}, {Das}, {Devlin}, {Dunkley},
  {D{\"u}nner}, {Farhang}, {Gralla}, {Hajian}, {Halpern}, {Hasselfield},
  {Hincks}, {Irwin}, {Kosowsky}, {Louis}, {Marriage}, {Moodley}, {Newburgh},
  {Niemack}, {Nolta}, {Page}, {Sehgal}, {Sherwin}, {Sievers}, {Sif{\'o}n},
  {Spergel}, {Staggs}, {Switzer}, \& {Wollack}}]{ACT2013}
{Calabrese}, E., {Hlozek}, R.~A., {Battaglia}, N., {et~al.} 2013, \prd, 87,
  103012

\bibitem[{{Copi} {et~al.}(2004){Copi}, {Huterer}, \& {Starkman}}]{Copi2004}
{Copi}, C.~J., {Huterer}, D., \& {Starkman}, G.~D. 2004, \prd, 70, 043515

\bibitem[{{Cornish} {et~al.}(2004){Cornish}, {Spergel}, {Starkman}, \&
  {Komatsu}}]{Cornish2004}
{Cornish}, N.~J., {Spergel}, D.~N., {Starkman}, G.~D., \& {Komatsu}, E. 2004,
  Physical Review Letters, 92, 201302

\bibitem[{{Crittenden} {et~al.}(1993){Crittenden}, {Davis}, \&
  {Steinhardt}}]{PolCom1993}
{Crittenden}, R., {Davis}, R.~L., \& {Steinhardt}, P.~J. 1993, \apjl, 417, L13

\bibitem[{{Crittenden} {et~al.}(1995){Crittenden}, {Coulson}, \&
  {Turok}}]{PolCom1995}
{Crittenden}, R.~G., {Coulson}, D., \& {Turok}, N.~G. 1995, \prd, 52, R5402

\bibitem[{{Delabrouille} {et~al.}(2009){Delabrouille}, {Cardoso}, {Le Jeune},
  {Betoule}, {Fay}, \& {Guilloux}}]{Delabrouille2009}
{Delabrouille}, J., {Cardoso}, J.-F., {Le Jeune}, M., {et~al.} 2009, \aap, 493,
  835

\bibitem[{{Eriksen} {et~al.}(2007{\natexlab{a}}){Eriksen}, {Banday},
  {G{\'o}rski}, {Hansen}, \& {Lilje}}]{Eriksen2007}
{Eriksen}, H.~K., {Banday}, A.~J., {G{\'o}rski}, K.~M., {Hansen}, F.~K., \&
  {Lilje}, P.~B. 2007{\natexlab{a}}, \apjl, 660, L81

\bibitem[{{Eriksen} {et~al.}(2008{\natexlab{a}}){Eriksen}, {Dickinson},
  {Jewell}, {Banday}, {G{\'o}rski}, \& {Lawrence}}]{Eriksen2008a}
{Eriksen}, H.~K., {Dickinson}, C., {Jewell}, J.~B., {et~al.}
  2008{\natexlab{a}}, \apjl, 672, L87

\bibitem[{{Eriksen} {et~al.}(2004){Eriksen}, {Hansen}, {Banday}, {G{\'o}rski},
  \& {Lilje}}]{Eriksen2004}
{Eriksen}, H.~K., {Hansen}, F.~K., {Banday}, A.~J., {G{\'o}rski}, K.~M., \&
  {Lilje}, P.~B. 2004, \apj, 605, 14

\bibitem[{{Eriksen} {et~al.}(2008{\natexlab{b}}){Eriksen}, {Jewell},
  {Dickinson}, {Banday}, {G{\'o}rski}, \& {Lawrence}}]{Eriksen2008b}
{Eriksen}, H.~K., {Jewell}, J.~B., {Dickinson}, C., {et~al.}
  2008{\natexlab{b}}, \apj, 676, 10

\bibitem[{{Eriksen} {et~al.}(2007{\natexlab{b}}){Eriksen}, {Huey}, {Saha},
  {Hansen}, {Dick}, {Banday}, {G{\'o}rski}, {Jain}, {Jewell}, {Knox}, {Larson},
  {O'Dwyer}, {Souradeep}, \& {Wandelt}}]{EriksenWMAP2007}
{Eriksen}, H.~K., {Huey}, G., {Saha}, R., {et~al.} 2007{\natexlab{b}}, \apj,
  656, 641

\bibitem[{{Gold} {et~al.}(2009){Gold}, {Bennett}, {Hill}, {Hinshaw}, {Odegard},
  {Page}, {Spergel}, {Weiland}, {Dunkley}, {Halpern}, {Jarosik}, {Kogut},
  {Komatsu}, {Larson}, {Meyer}, {Nolta}, {Wollack}, \& {Wright}}]{Gold2009}
{Gold}, B., {Bennett}, C.~L., {Hill}, R.~S., {et~al.} 2009, \apjs, 180, 265

\bibitem[{{Gold} {et~al.}(2011){Gold}, {Odegard}, {Weiland}, {Hill}, {Kogut},
  {Bennett}, {Hinshaw}, {Chen}, {Dunkley}, {Halpern}, {Jarosik}, {Komatsu},
  {Larson}, {Limon}, {Meyer}, {Nolta}, {Page}, {Smith}, {Spergel}, {Tucker},
  {Wollack}, \& {Wright}}]{Gold2011}
{Gold}, B., {Odegard}, N., {Weiland}, J.~L., {et~al.} 2011, \apjs, 192, 15

\bibitem[{{G{\'o}rski} {et~al.}(2005){G{\'o}rski}, {Hivon}, {Banday},
  {Wandelt}, {Hansen}, {Reinecke}, \& {Bartelmann}}]{Gorski2005}
{G{\'o}rski}, K.~M., {Hivon}, E., {Banday}, A.~J., {et~al.} 2005, \apj, 622,
  759

\bibitem[{{Guth}(1981)}]{GuthInflation1981}
{Guth}, A.~H. 1981, \prd, 23, 347

\bibitem[{{Hajian} \& {Souradeep}(2003)}]{HajianSouradeep2003}
{Hajian}, A., \& {Souradeep}, T. 2003, \apjl, 597, L5

\bibitem[{{Hajian} \& {Souradeep}(2006)}]{HajianSouradeep2006}
---. 2006, \prd, 74, 123521

\bibitem[{{Hajian} {et~al.}(2005){Hajian}, {Souradeep}, \&
  {Cornish}}]{HajianSouradeep2005}
{Hajian}, A., {Souradeep}, T., \& {Cornish}, N. 2005, \apjl, 618, L63

\bibitem[{{Hinshaw} {et~al.}(2007){Hinshaw}, {Nolta}, {Bennett}, {Bean},
  {Dore}, {Greason}, {Halpern}, {Hill}, {Jarosik}, {Kogut}, {Komatsu}, {Limon},
  {Odegard}, {Meyer}, {Page}, {Peiris}, {Spergel}, {Tucker}, {Verde},
  {Weiland}, {Wollack}, \& {Wright}}]{Hinshaw_07}
{Hinshaw}, G., {Nolta}, M.~R., {Bennett}, C.~L., {et~al.} 2007, Astrophys. J.
  Suppl. Ser., 170, 288

\bibitem[{{Hinshaw} {et~al.}(2013){Hinshaw}, {Larson}, {Komatsu}, {Spergel},
  {Bennett}, {Dunkley}, {Nolta}, {Halpern}, {Hill}, {Odegard}, {Page}, {Smith},
  {Weiland}, {Gold}, {Jarosik}, {Kogut}, {Limon}, {Meyer}, {Tucker}, {Wollack},
  \& {Wright}}]{Hinshaw2013}
{Hinshaw}, G., {Larson}, D., {Komatsu}, E., {et~al.} 2013, \apjs, 208, 19

\bibitem[{{Hivon} {et~al.}(2002){Hivon}, {G{\'o}rski}, {Netterfield}, {Crill},
  {Prunet}, \& {Hansen}}]{Hivon2002}
{Hivon}, E., {G{\'o}rski}, K.~M., {Netterfield}, C.~B., {et~al.} 2002, \apj,
  567, 2

\bibitem[{{Hou} {et~al.}(2014){Hou}, {Reichardt}, {Story}, {Follin}, {Keisler},
  {Aird}, {Benson}, {Bleem}, {Carlstrom}, {Chang}, {Cho}, {Crawford}, {Crites},
  {de Haan}, {de Putter}, {Dobbs}, {Dodelson}, {Dudley}, {George}, {Halverson},
  {Holder}, {Holzapfel}, {Hoover}, {Hrubes}, {Joy}, {Knox}, {Lee}, {Leitch},
  {Lueker}, {Luong-Van}, {McMahon}, {Mehl}, {Meyer}, {Millea}, {Mohr},
  {Montroy}, {Padin}, {Plagge}, {Pryke}, {Ruhl}, {Sayre}, {Schaffer}, {Shaw},
  {Shirokoff}, {Spieler}, {Staniszewski}, {Stark}, {van Engelen},
  {Vanderlinde}, {Vieira}, {Williamson}, \& {Zahn}}]{SPT2_2014}
{Hou}, Z., {Reichardt}, C.~L., {Story}, K.~T., {et~al.} 2014, \apj, 782, 74

\bibitem[{{Lachieze-Rey} \& {Luminet}(1995)}]{Lachieze1995}
{Lachieze-Rey}, M., \& {Luminet}, J. 1995, \physrep, 254, 135

\bibitem[{{Levin}(2002)}]{Levin2002}
{Levin}, J. 2002, \physrep, 365, 251

\bibitem[{{Linde}(1983)}]{LindeInflation1983}
{Linde}, A.~D. 1983, Physics Letters B, 129, 177

\bibitem[{{Luminet}(2016)}]{Luminet2016}
{Luminet}, J.-P. 2016, Universe, 2, 1

\bibitem[{{Peiris} {et~al.}(2003){Peiris}, {Komatsu}, {Verde}, {Spergel},
  {Bennett}, {Halpern}, {Hinshaw}, {Jarosik}, {Kogut}, {Limon}, {Meyer},
  {Page}, {Tucker}, {Wollack}, \& {Wright}}]{WMAPInflation2003}
{Peiris}, H.~V., {Komatsu}, E., {Verde}, L., {et~al.} 2003, \apjs, 148, 213

\bibitem[{{Planck Collaboration} {et~al.}(2015){Planck Collaboration}, {Adam},
  {Ade}, {Aghanim}, {Alves}, {Arnaud}, {Ashdown}, {Aumont}, {Baccigalupi},
  {Banday}, \& et~al.}]{Planck2015_fg}
{Planck Collaboration}, {Adam}, R., {Ade}, P.~A.~R., {et~al.} 2015, ArXiv
  e-prints, arXiv:1502.01588

\bibitem[{{Planck Collaboration} {et~al.}(2016{\natexlab{a}}){Planck
  Collaboration}, {Adam}, {Ade}, {Aghanim}, {Akrami}, {Alves}, {Arg{\"u}eso},
  {Arnaud}, {Arroja}, {Ashdown}, \& et~al.}]{Planck2016}
---. 2016{\natexlab{a}}, \aap, 594, A1

\bibitem[{{Planck Collaboration} {et~al.}(2016{\natexlab{b}}){Planck
  Collaboration}, {Ade}, {Aghanim}, {Ashdown}, {Aumont}, {Baccigalupi},
  {Ballardini}, {Banday}, {Barreiro}, {Bartolo}, \& et~al.}]{Planck2016lfi}
{Planck Collaboration}, {Ade}, P.~A.~R., {Aghanim}, N., {et~al.}
  2016{\natexlab{b}}, \aap, 594, A2

\bibitem[{{Planck Collaboration} {et~al.}(2016{\natexlab{c}}){Planck
  Collaboration}, {Adam}, {Ade}, {Aghanim}, {Arnaud}, {Ashdown}, {Aumont},
  {Baccigalupi}, {Banday}, {Barreiro}, \& et~al.}]{Planck2016_CMB}
{Planck Collaboration}, {Adam}, R., {Ade}, P.~A.~R., {et~al.}
  2016{\natexlab{c}}, \aap, 594, A9

\bibitem[{{Planck Collaboration} {et~al.}(2016{\natexlab{d}}){Planck
  Collaboration}, {Adam}, {Ade}, {Aghanim}, {Arnaud}, {Ashdown}, {Aumont},
  {Baccigalupi}, {Banday}, {Barreiro}, \& et~al.}]{Planck2016hfi}
---. 2016{\natexlab{d}}, \aap, 594, A8

\bibitem[{{Planck Collaboration} {et~al.}(2016{\natexlab{e}}){Planck
  Collaboration}, {Ade}, {Aghanim}, {Arnaud}, {Ashdown}, {Aumont},
  {Baccigalupi}, {Banday}, {Barreiro}, {Bartlett}, \&
  et~al.}]{PlanckCosmoParam2016}
{Planck Collaboration}, {Ade}, P.~A.~R., {Aghanim}, N., {et~al.}
  2016{\natexlab{e}}, \aap, 594, A13

\bibitem[{{Planck Collaboration} {et~al.}(2016{\natexlab{f}}){Planck
  Collaboration}, {Ade}, {Aghanim}, {Akrami}, {Aluri}, {Arnaud}, {Ashdown},
  {Aumont}, {Baccigalupi}, {Banday}, \& et~al.}]{PlanckIso2016}
---. 2016{\natexlab{f}}, \aap, 594, A16

\bibitem[{{Planck Collaboration} {et~al.}(2016{\natexlab{g}}){Planck
  Collaboration}, {Ade}, {Aghanim}, {Arnaud}, {Ashdown}, {Aumont},
  {Baccigalupi}, {Banday}, {Barreiro}, {Bartolo}, \&
  et~al.}]{PlanckTopology2016}
---. 2016{\natexlab{g}}, \aap, 594, A18

\bibitem[{{Planck Collaboration} {et~al.}(2016{\natexlab{h}}){Planck
  Collaboration}, {Ade}, {Aghanim}, {Arnaud}, {Arroja}, {Ashdown}, {Aumont},
  {Baccigalupi}, {Ballardini}, {Banday}, \& et~al.}]{PlanckInflation2016}
---. 2016{\natexlab{h}}, \aap, 594, A20

\bibitem[{{Saha}(2011)}]{Saha2011}
{Saha}, R. 2011, \apjl, 739, L56

\bibitem[{{Saha} \& {Aluri}(2016)}]{Saha2016}
{Saha}, R., \& {Aluri}, P.~K. 2016, \apj, 829, 113

\bibitem[{{Saha} {et~al.}(2006){Saha}, {Jain}, \& {Souradeep}}]{Saha2006}
{Saha}, R., {Jain}, P., \& {Souradeep}, T. 2006, \apjl, 645, L89

\bibitem[{{Saha} {et~al.}(2008){Saha}, {Prunet}, {Jain}, \&
  {Souradeep}}]{Saha2008}
{Saha}, R., {Prunet}, S., {Jain}, P., \& {Souradeep}, T. 2008, \prd, 78, 023003

\bibitem[{{Samal} {et~al.}(2008){Samal}, {Saha}, {Jain}, \&
  {Ralston}}]{Samal2008}
{Samal}, P.~K., {Saha}, R., {Jain}, P., \& {Ralston}, J.~P. 2008, \mnras, 385,
  1718

\bibitem[{{Spergel} \& {Zaldarriaga}(1997)}]{PolCom1997}
{Spergel}, D.~N., \& {Zaldarriaga}, M. 1997, Physical Review Letters, 79, 2180

\bibitem[{{Starobinsky}(1982)}]{StarobinskyInflation1982}
{Starobinsky}, A.~A. 1982, Physics Letters B, 117, 175

\bibitem[{{Story} {et~al.}(2013){Story}, {Reichardt}, {Hou}, {Keisler}, {Aird},
  {Benson}, {Bleem}, {Carlstrom}, {Chang}, {Cho}, {Crawford}, {Crites}, {de
  Haan}, {Dobbs}, {Dudley}, {Follin}, {George}, {Halverson}, {Holder},
  {Holzapfel}, {Hoover}, {Hrubes}, {Joy}, {Knox}, {Lee}, {Leitch}, {Lueker},
  {Luong-Van}, {McMahon}, {Mehl}, {Meyer}, {Millea}, {Mohr}, {Montroy},
  {Padin}, {Plagge}, {Pryke}, {Ruhl}, {Sayre}, {Schaffer}, {Shaw}, {Shirokoff},
  {Spieler}, {Staniszewski}, {Stark}, {van Engelen}, {Vanderlinde}, {Vieira},
  {Williamson}, \& {Zahn}}]{SPT1_2013}
{Story}, K.~T., {Reichardt}, C.~L., {Hou}, Z., {et~al.} 2013, \apj, 779, 86

\bibitem[{{Tegmark} {et~al.}(2003){Tegmark}, {de Oliveira-Costa}, \&
  {Hamilton}}]{Tegmark2003}
{Tegmark}, M., {de Oliveira-Costa}, A., \& {Hamilton}, A.~J. 2003, Phys. Rev.
  D, 68, 123523

\bibitem[{{Tegmark} \& {Efstathiou}(1996)}]{Tegmark96}
{Tegmark}, M., \& {Efstathiou}, G. 1996, Mon. Not. R. Astron. Soc., 281, 1297

\end{thebibliography}

\end{document}